\documentclass[aps,a4paper,twocolumn]{revtex4-2}
\usepackage{amsmath}
\usepackage{graphicx}
\usepackage{subfigure}
\usepackage[usenames,dvipsnames]{color}
\definecolor{darkblue}{RGB}{0,0,196}
\usepackage[colorlinks=true,linkcolor=darkblue,citecolor=darkblue,urlcolor=darkblue]{hyperref}
\usepackage{setspace}
\usepackage{hyperref}
\usepackage{xcolor}
\hypersetup{
  colorlinks   = true, 
  urlcolor     = blue, 
  linkcolor    = blue, 
  citecolor   = blue 
}
\usepackage{graphicx}
\usepackage{amsmath,bbm}
\usepackage{amssymb,bm}
\usepackage{yfonts}
\usepackage{comment}

\usepackage{dcolumn}
\usepackage{ulem}
\usepackage{amsmath,amssymb,amsfonts,bm}

\begin{document}

\title{Role of clustered nuclear geometry in particle production through \\ p--C and p--O collisions at the Large Hadron Collider}

\author{Aswathy Menon K R$^{1}$}\email[]{Aswathy.Menon@cern.ch}
\author{Suraj Prasad$^{1}$}\email[]{Suraj.Prasad@cern.ch}
\author{Neelkamal Mallick$^{1}$}\email[]{Neelkamal.Mallick@cern.ch}
\author{Raghunath Sahoo$^{1}$}\email[Corresponding author: ]{Raghunath.Sahoo@cern.ch}
\affiliation{$^{1}$Department of Physics, Indian Institute of Technology Indore, Simrol, Indore 453552, India}

\date{\today}

\begin{abstract}

Long-range multi-particle correlations in heavy-ion collisions have shown conclusive evidence of the hydrodynamic behavior of strongly interacting matter and are associated with the final-state azimuthal momentum anisotropy. In small collision systems, azimuthal anisotropy can be influenced by the hadronization mechanism and residual jet-like correlations. Thus, one of the motives of the planned p--O and O--O collisions at the LHC and RHIC is to understand the origin of small system collectivity. As the anisotropic flow coefficients ($v_n$) are sensitive to the initial-state effects including nuclear shape, deformation, and charge density profiles, studies involving $^{12}$C and $^{16}$O nuclei are transpiring due to the presence of exotic $\alpha$ ($^{4}$He) clusters in such nuclei. In this study, for the first time, we investigate the effects of nuclear $\alpha$--clusters on the azimuthal anisotropy of the final-state hadrons in p--C and p--O collisions at $\sqrt{s_{\rm NN}}= 9.9$~TeV within a multi-phase transport model framework. We report the transverse momentum ($p_{\rm T}$) and pseudorapidity ($\eta$) spectra, participant eccentricity ($\epsilon_2$) and triangularity ($\epsilon_3$), and estimate the elliptic flow ($v_2$) and triangular flow ($v_3$) of the final-state hadrons using the two-particle cumulant method. These results are compared with a model-independent Sum of Gaussians (SOG) type nuclear density profile for $^{12}$C and $^{16}$O nuclei.

\end{abstract}

\maketitle


\section{Introduction}
\label{sec1}
The world's most powerful particle accelerators, the Relativistic Heavy-Ion Collider (RHIC) at BNL, USA, and the Large Hadron Collider (LHC) at CERN, Switzerland, offer unparalleled experimental capabilities to perform high-energy hadronic and nuclear collisions to study the behavior of nuclear matter under extreme conditions of temperature and energy density. Such ultra-relativistic heavy-ion collisions produce a state of locally thermalized and deconfined partonic medium, known as the quark-gluon plasma (QGP). While the existence of QGP has long been established in heavy-ion collisions, its presence in small collision systems is a matter of intense research, recently. Thanks to the recent measurements of heavy-ion-like features in high multiplicity pp collisions such as strangeness enhancement~\cite{ALICE:2016fzo}, ridge-like structure~\cite{CMS:2015fgy, ALICE:2013snk}, and radial flow effects~\cite{ALICE:2013wgn, ALICE:2016dei, CMS:2016fnw}, the possible existence of QGP droplets in small collision systems needs to be re-examined. As p--C and p--O collisions fill the multiplicity gap between pp to p--Pb and peripheral Pb--Pb collisions, they provide a perfect system size for studying several heavy-ion-like effects in small collision systems at the RHIC and LHC energies~\cite{Brewer:2021kiv, Katz:2019qwv}. 

\begin{figure}
\includegraphics[scale=0.25]{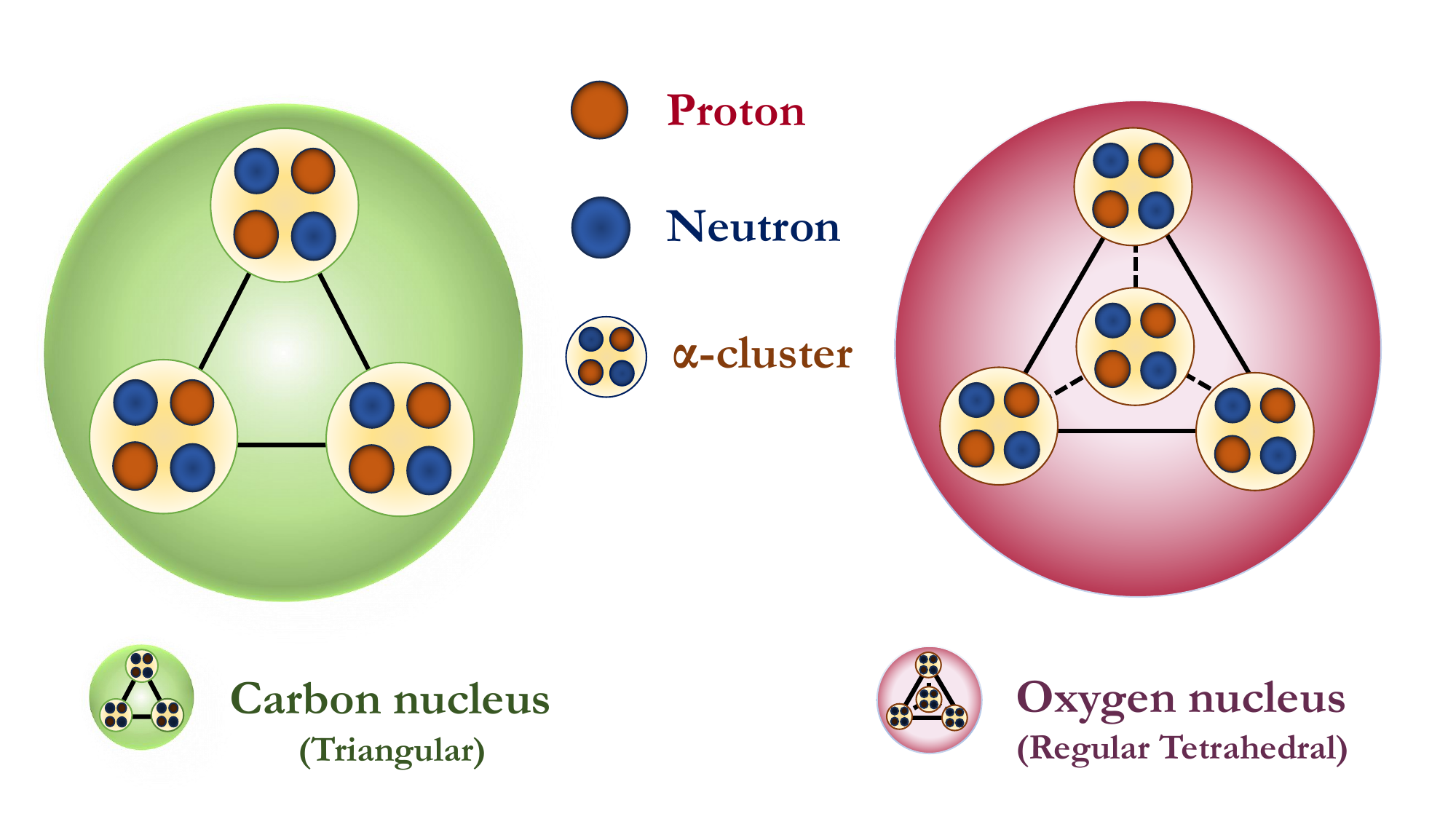}
\caption{(Color online) Pictorial representation of the arrangement of $\alpha$--particles inside $^{12}$C (left) and $^{16}$O (right) nucleus.}
\label{fig:ALPHApOpC}
\end{figure}

In non-central heavy-ion collisions, the appearance of strong final-state azimuthal momentum anisotropy is associated with the hydrodynamic collective expansion of the medium~\cite{Heinz:2013th, Ollitrault:1992bk}. This collectivity is usually quantified as the Fourier coefficients ($v_n$) of the final-state azimuthal momentum distribution ($dN/d\phi$)~\cite{Voloshin:2008dg, Voloshin:1994mz}. Azimuthal anisotropy mainly arises from the initial nuclear geometry and fluctuations in the energy and entropy density, which further gets embedded in the system through its equation-of-state and transport coefficients~\cite{Heinz:2013th, Ollitrault:1992bk}. Hydrodynamic model predictions when confronted with the experimental findings have shown conclusive evidence of the collective expansion of the thermalized medium and strongly hint at the formation of QGP in heavy-ion collisions~\cite{STAR:2005gfr}. However, such collective behavior is usually not anticipated in small collision systems, thus, the applicability of hydrodynamics, and hence, thermalization in the early stages of the collision is still debatable in small collision systems.

The anisotropic flow coefficients, such as elliptic flow ($v_{2}$) and triangular flow ($v_{3}$), are found to be sensitive to the initial nuclear distribution, nuclear shape deformation, and fluctuations in the nuclear overlap region~\cite{Behera:2023nwj, Giacalone:2021udy, Haque:2019vgi}. In Xe--Xe collisions at the LHC, the presence of a deformed nuclear structure gives rise to a higher value of elliptic flow in the most central collisions compared to Pb--Pb collisions~\cite{ALICE:2021ibz, ALICE:2018lao, ATLAS:2019dct, CMS:2019cyz}. In addition, a larger density fluctuation due to a smaller number of participating nucleons in Xe--Xe collisions gives rise to a larger value of triangular flow in contrast to Pb--Pb collisions~\cite{ALICE:2021ibz, ALICE:2018lao}. A similar study using a multi-phase transport model (AMPT) in U--U collisions at the RHIC energy shows that the values of anisotropic flow coefficients change with a change in the deformation parameter of the nuclear distribution~\cite{Giacalone:2021udy}. However, the impact of this deformation is larger for elliptic flow in contrast to the triangular flow~\cite{Giacalone:2021udy, Haque:2019vgi, PHENIX:2018lia}. Additionally, recent observations of ridge-like structures in p--Pb collisions have brought significant attention to asymmetric collisions, which are customarily considered to provide a baseline measurement for the cold nuclear effects~\cite{CMS:2012qk, ALICE:2012eyl, ATLAS:2012cix, CMS:2013jlh, CMS:2016fnw}, where the presence of strong long-range multi-particle correlations are not expected. Further, the measured values of anisotropic flow coefficients in p--Pb collisions are comparable to that of peripheral Pb--Pb collisions~\cite{ALICE:2014dwt, ALICE:2019zfl, Tang:2023wcd}. Although, in p--Pb collisions, the ordering of the magnitude of flow coefficients, i.e., $v_{2}>v_{3}>v_{4}$, is not as strong as in Pb--Pb collisions, the bare presence of signatures of collectivity in p--Pb collisions makes the study of asymmetric collisions a vital contribution of the heavy-ion physics community. 

Nuclei having $4n$ number of nucleons, are theorized to have $\alpha$--clusters ($^{4}$He-nucleus), which include $^8$Be, $^{12}$C, and $^{16}$O nuclei. In $^{12}$C, the $\alpha$--particles arrange themselves at the corners of an equilateral triangle. Similarly, nucleons inside $^{16}$O are expected to configure themselves in four $\alpha$--clusters forming a regular tetrahedral geometry~\cite{gamow, Wheeler:1937zza, Bijker:2014tka, Wang:2019dpl, He:2014iqa, He:2021uko}. Figure~\ref{fig:ALPHApOpC} shows the pictorial representation of the arrangement of nucleons inside an $\alpha$--particle, and the arrangement of $\alpha$--clusters inside $^{12}$C (left) and $^{16}$O (right) nuclei. Studies involving $^{12}$C and $^{16}$O nuclei have the potential to explore the emergence of collectivity in small systems in addition to the study of $\alpha$--cluster nuclear geometry effects on the final-state observables. Additionally, studies involving forward kinematics in p--O collisions are crucial in understanding the interaction of cosmic rays (mostly protons) with the Earth's atmosphere (mostly $^{14}$N and $^{16}$O). It could also help resolve the outstanding muon puzzle and perform precise measurements of the $\pi^0$ energy fraction ($R = E_{\pi^0}/E_{\rm all~hadrons}$), a smaller value of which is potentially attributed to strangeness enhancement, and hence hints towards the formation of QGP droplets~\cite{Brewer:2021kiv}.

With all these tempting motivations, in recent years, several studies have been performed involving collisions of $^{12}$C, $^{16}$O and  $^{20}$Ne nuclei~\cite{Bozek:2014cva, Broniowski:2013dia, Li:2020vrg, Rybczynski:2019adt, Sievert:2019zjr, Huang:2019tgz, Behera:2023nwj, Behera:2021zhi, YuanyuanWang:2024sgp, Zhang:2024vkh, Giacalone:2024luz}. It includes the studies based on different hydrodynamic models~\cite{Lim:2018huo, Summerfield:2021oex, Schenke:2020mbo, YuanyuanWang:2024sgp}, Glauber Monte Carlo~\cite{Rybczynski:2019adt, Sievert:2019zjr, Huang:2019tgz}, parton energy loss~\cite{Huss:2020whe}, and jet quenching effects~\cite{Zakharov:2021uza}. Some of these studies also investigate the possible signatures of $\alpha$--clusters in $^{12}$C and $^{16}$O nuclei by studying the final-state particle production and anisotropic flow measurements~\cite{Bozek:2014cva, Broniowski:2013dia, Li:2020vrg, Ding:2023ibq, YuanyuanWang:2024sgp, Wang:2021ghq, Rybczynski:2017nrx, alpha-frag, Behera:2023nwj, Behera:2023oxe}. In Ref.~\cite{Bozek:2014cva}, an enhanced value of triangular flow in C--Au collisions is observed for an $\alpha$--clustered $^{12}$C nucleus. In Ref.~\cite{Behera:2023nwj}, authors show that the presence of an $\alpha$--clustered structure in $^{16}$O nucleus leads to a high value of triangular flow in most central O--O collisions, which are well observed in the ratio $v_{3}/v_{2}$. In addition, an away-side broadening in the two-particle correlation function is also reported for O--O collisions with $\alpha$--clusters as compared to the Woods-Saxon nuclear profile~\cite{Behera:2023nwj}. Although there have been a few studies that aim to establish a clear signature of the presence of $\alpha$--clusters in $^{16}$O and $^{12}$C nucleus, the choice of p--O and p--C is novel to this study. In addition, the studies of anisotropic flow in asymmetric collision systems like p--O and p--C can accord the observations of collectivity in p--Pb collisions, which can be an interesting and a new addition to the existing results of the geometry scan program at RHIC~\cite{PHENIX:2016cfs, OrjuelaKoop:2015jss, PHENIX:2022nht, PHENIX:2021ubk}.

In this study, we incorporate an $\alpha$--cluster-type initial geometry for $^{12}$C and $^{16}$O nuclei, and report the first measurements of transverse momentum ($p_{\rm T}$) and pseudorapidity ($\eta$) spectra, participant eccentricity ($\epsilon_2$) and triangularity ($\epsilon_3$), and elliptic flow ($v_2$) and triangular flow ($v_3$) of the final-state hadrons, and their scaling in p--O and p--C collisions at $\sqrt{s_{\rm NN}}=9.9$ TeV using AMPT. We use the two-particle cumulant method with a relative pseudorapidity cut to minimize the effects of non-flow correlations. These results are compared with a model-independent Sum of Gaussians (SOG) type nuclear density profile for $^{12}$C and $^{16}$O nuclei as an unclustered description of nuclei is essential for a good understanding of the effects in the final state.

The paper is organized as follows. We start with a brief introduction and motivation of the study in Section~\ref{sec1} followed by a discussion on the implementation of various nuclear density profiles in Section~\ref{sec2}. In Section~\ref{sec3}, we present and discuss the results. Finally, in Section~\ref{sec4}, we summarize the study with important findings and present a brief outlook for the future. A detailed description of the event generation using AMPT and the anisotropic flow estimation using the two-particle Q-cumulant method are discussed in Appendix~\ref{Appendix-I}.

\section{NUCLEAR DENSITY PROFILE IMPLEMENTATION}
\label{sec2}
This study implements two different nuclear density profiles for the $^{12}$C and $^{16}$O nuclei: $\alpha$--cluster type geometrical distribution and a model-independent Sum of Gaussians type nuclear density profile. The technical details regarding the implementation of the two density profiles are discussed below. 

\subsubsection{\textbf{Sum Of Gaussians (SOG) density profile}}
To represent the nuclear charge density and extract charge density parameters in a model-independent fashion as approximated in experiments, one fits the nuclear densities with a sum of Gaussian functions (SOG) \cite{Shukla:2011, deVries:1987}. SOG can accurately represent complex nuclear densities by exploiting the parameters of individual Gaussian functions, which makes SOG applicable to a broad list of nuclei. SOG can produce smooth and continuous form of the nuclear density profiles, consequently simplifying many theoretical calculations due to the well-known properties of the Gaussian distribution function. In addition, SOG can produce a more realistic and precise description of the nuclear profiles as compared to the traditional uniform density or a Woods-Saxon description of the nucleus~\cite{Shukla:2011}. In this study, we use the sum of two Gaussian functions to simulate $^{12}$C and $^{16}$O nuclear profiles. The SOG nuclear charge density ($\rho(r)$) as a function of radial distance ($r$) is expressed as,
 \begin{equation}
       \rho (r) = C_{1}e^{-a_{1}r^{2}} + C_{2}e^{-a_{2}r^{2}}.
\end{equation}
Here, the coefficients $C_{1}$, $C_{2}$, $a_{1}$, and $a_{2}$ for the respective nuclei are obtained from Ref.~\cite{Shukla:2011}. Due to the rapid decrease of the Gaussian tail, the values of the charge density at various radii ($\rho(r)$) are advantageously decoupled. Table~\ref{tab:SOGpars} shows the parameters of SOG that have been considered in this study to simulate the $^{12}$C and $^{16}$O nuclei.

\begin{table}
    \centering
    \caption{Parameters of Sum of Gaussians nuclear density profile chosen for $^{12}$C and $^{16}$O nucleus~\cite{Shukla:2011}.}
    \label{tab:SOGpars}
    \begin{tabular}{|c|c|c|c|c|}
\hline 
 Nucleus & $C_1$ & $C_2$ & $a_1$ & $a_2$ \\ \hline \hline
 Carbon ($^{12}$C) & -0.162 & 0.340 & 0.554 &  0.280 \\ \hline
 Oxygen ($^{16}$O) & -0.539 & 0.729 & 0.407 & 0.300 \\ \hline
    \end{tabular}
\end{table}

\subsubsection{\textbf{\texorpdfstring{$\alpha$--clustered nuclear geometry}{}}}

The $^4$He nucleus consisting of a pair of protons and a pair of neutrons is known as the $\alpha$--particle. As discussed earlier, nucleons of certain light nuclei can cluster into groups of $\alpha$--particles forming a stable geometrical shape. For this, the nuclei should have $4n$ ($n$ is a positive integer) number of nucleons. In $^{12}$C, three such $\alpha$--particles cluster into forming an equilateral triangle, whereas in $^{16}$O, four $\alpha$--particles cluster into a regular tetrahedral arrangement. This $\alpha$--cluster geometry is believed to account for the additional stability of the nucleus. In our study, for the first time, we have implemented such $\alpha$--cluster geometry for $^{12}$C and $^{16}$O nuclei to study the initial-state effects in particle production through p--C and p--O collisions at the LHC. The technical details of the implementation of $\alpha$--cluster structure for $^{16}$O and $^{12}$C nuclei are discussed below.

\begin{itemize}

\item \textbf{Tetrahedral geometry for $^{16}$O}
  
For the $^{16}\rm O$ nucleus, four $\alpha$--particles are situated at the four vertices of a regular tetrahedron with side length 3.42~fm, which makes the root mean squared (rms) radius of $^{16}\rm O$ to be 2.699~fm~\cite{Li:2020vrg, Behera:2023nwj, Behera:2021zhi}. One can visualize the nucleons inside the $^{16}$O nuclei as depicted in Fig.~\ref{fig:ALPHApOpC}. Nucleons inside the $\alpha$--particle are sampled using the following Woods-Saxon density profile in terms of a
three-parameter Fermi (3pF) distribution,
\begin{equation}
\rho(r) = \frac{\rho_{0} \Big(1+ w \big(\frac{r}{r_{0}}\big)^{2}\Big)}{1 + \exp\big(\frac{r - r_{0}}{a}\big)}. 
\label{eq:WS}
\end{equation}
Here, $\rho(r)$ is the nuclear charge density at a radial distance $r$ from the center of the nucleus. $r_0$ refers to the mean radius of the nucleus, $w$ is the nuclear deformation parameter, and $a$ is the skin depth. The values of these parameters are chosen for the $^4$He nucleus as $r_{0}$ = 0.964 fm, $w$ = 0.517, and $a$ = 0.322 fm, which corresponds to the rms radius of 1.676~fm for the $\alpha$--particle. To include a finite volume effect for each nucleon, the minimum separation distance between any two nucleons is set to be 0.4~fm~\cite{Loizides:2017ack}. The spatial orientation of the tetrahedron is randomized before each collision for both the target and projectile nuclei.
 
\item \textbf{Triangular geometry for $^{12}$C}

Here, three $\alpha$--particles arrange themselves at the corners of an equilateral triangle of side length 3.10~fm leading to an rms radius of 2.47~fm for the $^{12}$C nucleus~\cite{Wang:2021ghq, Angeli:2013epw}. A depiction of the nucleons inside an $\alpha$-clustered $^{12}$C nuclei can be seen in Fig.~\ref{fig:ALPHApOpC}. The nucleons inside the $\alpha$--particles are sampled using a similar procedure as discussed above for the case of $^{16}\rm O$ nucleus.
 
\end{itemize}

\begin{table*} [!hpt]
                \centering 
                \caption{The impact parameter, average number of nucleon-nucleon binary collisions and average number of nucleon participants for different density profiles and in different centrality classes for p--O collisions at $\sqrt{s_{\rm{NN}}}$ = 9.9 TeV. Centrality selection is done through 
                geometrical slicing, {\it i.e.}, from the impact parameter obtained from the Glauber model~\cite{Loizides:2017ack}.}
                 \label{tab:glaubpO}
                  \scalebox{1.0}{
                \begin{tabular}{|c |c |c | c| c |c  |c  |c | c  |}
                \hline
                \textbf{p--O} & \multicolumn{4}{c|}{{\textbf{SOG}}} & \multicolumn{4}{c|}{{\textbf{$\alpha$--cluster}}}\\ 
                \hline
                \textbf{Centrality(\%)} & $b_{\rm{min}}$(fm) & $b_{\rm{max}}$(fm) & $\langle N_{\rm{coll}} \rangle$ $\pm$ rms & $\langle N_{\rm{part}} \rangle$  $\pm$  rms  & $b_{\rm{min}}$(fm) & $b_{\rm{max}}$(fm) & $\langle N_{\rm{coll}} \rangle$  $\pm$  rms  & $\langle N_{\rm{part}} \rangle$ $\pm$  rms  \\
                \hline \hline
                0-5 & 0 & 0.81  & 6.64 $\pm$ 2.12 & 7.64 $\pm$ 2.12 & 0 & 0.80 & 5.65 $\pm$ 1.64 & 6.65 $\pm$ 1.64 \\ \hline

5-10 & 0.81 & 1.15 &5.94 $\pm$ 1.92  & 6.94 $\pm$ 1.92 & 0.80 & 1.14 & 5.30 $\pm$ 1.64 & 6.30 $\pm$ 1.64 \\ \hline

10-20 & 1.15 & 1.63 & 5.00 $\pm$ 1.75 & 6.00 $\pm$ 1.75 & 1.14 & 1.61 & 4.78 $\pm$ 1.56 & 5.78 $\pm$ 1.56 \\ \hline

20-30 & 1.63 & 2.00 & 3.96 $\pm$ 1.56 & 4.96 $\pm$ 1.56 & 1.61 & 1.97 & 4.08 $\pm$ 1.40 & 5.08 $\pm$ 1.40 \\ \hline

30-40 & 2.00 &2.32  &3.16 $\pm$ 1.40  &  4.16 $\pm$ 1.40 & 1.97 & 2.28 & 3.43 $\pm$ 1.27 & 4.43 $\pm$ 1.27 \\ \hline

40-50 & 2.32 &2.60  & 2.54 $\pm$ 1.24 & 3.54 $\pm$ 1.24 & 2.28 & 2.56 & 2.84 $\pm$ 1.15 & 3.84 $\pm$ 1.15 \\ \hline

50-60 & 2.60 & 2.89 & 2.08 $\pm$ 1.06 & 3.08 $\pm$ 1.06 & 2.56 & 2.82 & 2.37 $\pm$ 1.03 & 3.37 $\pm$ 1.03 \\ \hline

60-70 & 2.89 & 3.18 & 1.73 $\pm$ 0.87 & 2.73 $\pm$ 0.87 & 2.82 & 3.08 & 1.99 $\pm$ 0.90 & 2.99 $\pm$ 0.90 \\ \hline

70-100 & 3.18 & 7.50 & 1.26 $\pm$ 0.54 & 2.26 $\pm$ 0.54 & 3.08 & 5.38 & 1.43 $\pm$ 0.65 & 2.43 $\pm$ 0.65 \\ \hline
                \end{tabular}     
                }

\end{table*}

\begin{table*} [!hpt]
                \centering 
                \caption{The impact parameter, average number of nucleon-nucleon binary collisions and average number of nucleon participants for different density profiles and in different centrality classes for p--C collisions at $\sqrt{s_{\rm{NN}}}$ = 9.9 TeV. Centrality selection is done through 
                geometrical slicing, {\it i.e.}, from the impact parameter obtained from the Glauber model.                 
                 \label{tab:glaubpC}}
                  \scalebox{1.0}{
                \begin{tabular}{|c |c |c | c| c |c  |c  |c | c  |}
                \hline
                \textbf{p--C} & \multicolumn{4}{c|}{{\textbf{SOG}}} & \multicolumn{4}{c|}{{\textbf{$\alpha$--cluster}}}\\ 
                \hline
                \textbf{Centrality(\%)} & $b_{\rm{min}}$(fm) & $b_{\rm{max}}$(fm) & $\langle N_{\rm{coll}} \rangle$ $\pm$ rms & $\langle N_{\rm{part}} \rangle$  $\pm$  rms  & $b_{\rm{min}}$(fm) & $b_{\rm{max}}$(fm) & $\langle N_{\rm{coll}} \rangle$  $\pm$  rms  & $\langle N_{\rm{part}} \rangle$ $\pm$  rms  \\
                \hline \hline
                0-5 & 0 & 0.76 & 5.49 $\pm$ 1.84 & 6.49 $\pm$ 1.84 & 0 & 0.71 & 6.05 $\pm$ 1.89 & 7.05 $\pm$ 1.89 \\ \hline
5-10 & 0.76 & 1.08 & 4.91 $\pm$ 1.69 & 5.91 $\pm$ 1.69 & 0.71 & 1.00 & 5.46 $\pm$ 1.68 & 6.46 $\pm$ 1.68 \\ \hline
10-20 & 1.08 & 1.53 & 4.14 $\pm$ 1.53 & 5.14 $\pm$ 1.53 & 1.00 & 1.42 & 4.70 $\pm$ 1.54 & 5.70 $\pm$ 1.54 \\ \hline
20-30 & 1.53 & 1.90 & 3.25 $\pm$ 1.34 & 4.25 $\pm$ 1.34 & 1.42 & 1.74 & 3.85 $\pm$ 1.38 & 4.85 $\pm$ 1.38 \\ \hline
30-40 & 1.90 & 2.18 & 2.61 $\pm$ 1.18 & 3.61 $\pm$ 1.18 & 1.74 & 2.02 & 3.14 $\pm$ 1.27 & 4.14 $\pm$ 1.27 \\ \hline
40-50 & 2.18 & 2.46 & 2.15 $\pm$ 1.03 & 3.15 $\pm$ 1.03 & 2.02 & 2.27 & 2.60 $\pm$ 1.15 & 3.60 $\pm$ 1.15 \\ \hline
50-60 & 2.46 & 2.74 & 1.80 $\pm$ 0.88 & 2.80 $\pm$ 0.88 & 2.27 & 2.51 & 2.18 $\pm$ 1.02 & 3.18 $\pm$ 1.02 \\ \hline
60-70 & 2.74 & 3.03 & 1.54 $\pm$ 0.73 & 2.54 $\pm$ 0.73 & 2.51 & 2.76 & 1.85 $\pm$ 0.90 & 2.85 $\pm$ 0.90 \\ \hline
70-100 & 3.03 & 7.44 & 1.19 $\pm$ 0.45 & 2.19 $\pm$ 0.45 & 2.76 & 5.86 & 1.36 $\pm$ 0.62 & 2.36 $\pm$ 0.62 \\ \hline
                \end{tabular}     
                }

\end{table*}

\section{Results and discussions}
\label{sec3}
\begin{figure*}[ht!]
\includegraphics[scale=0.35]{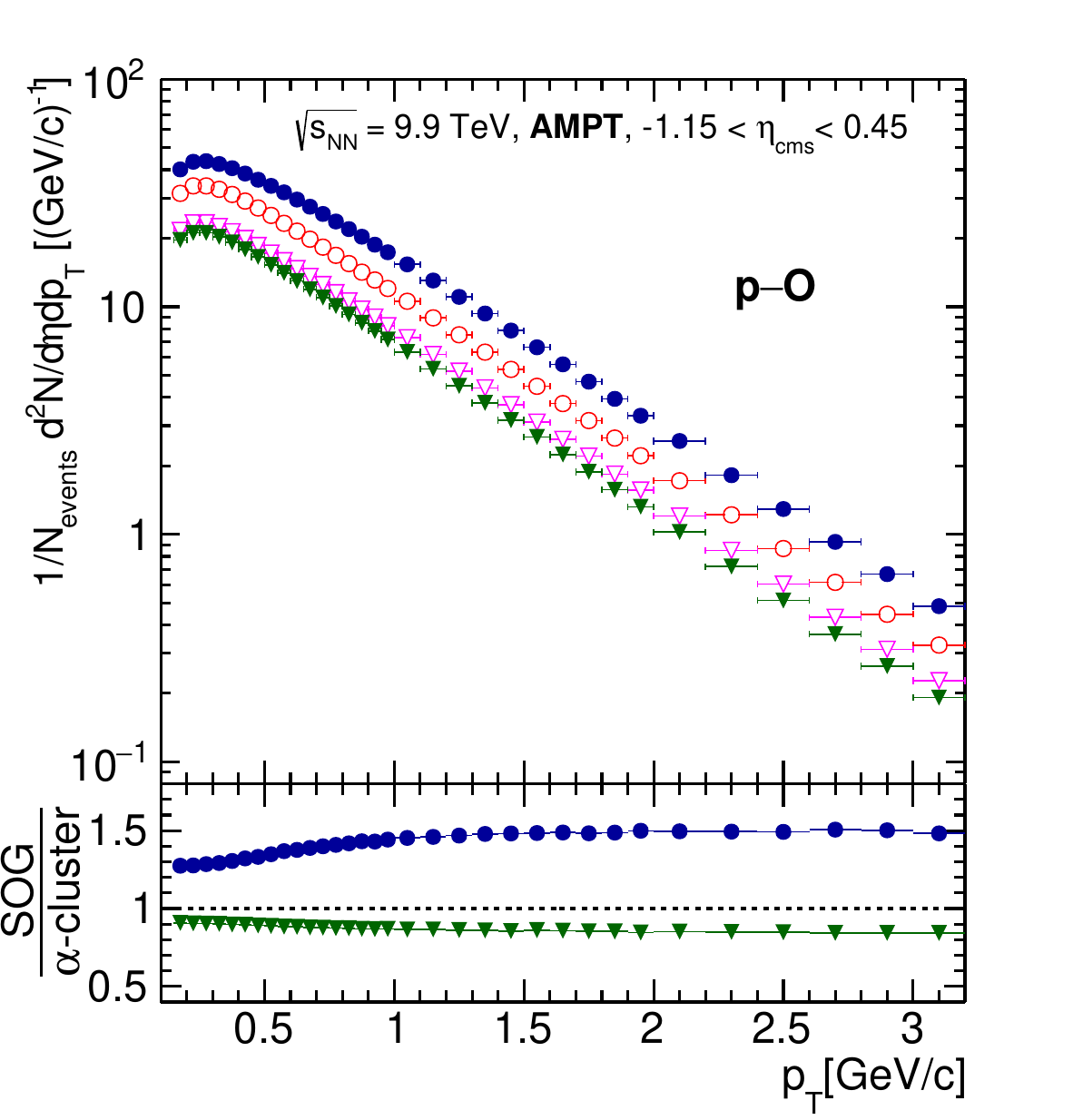}
\includegraphics[scale=0.35]{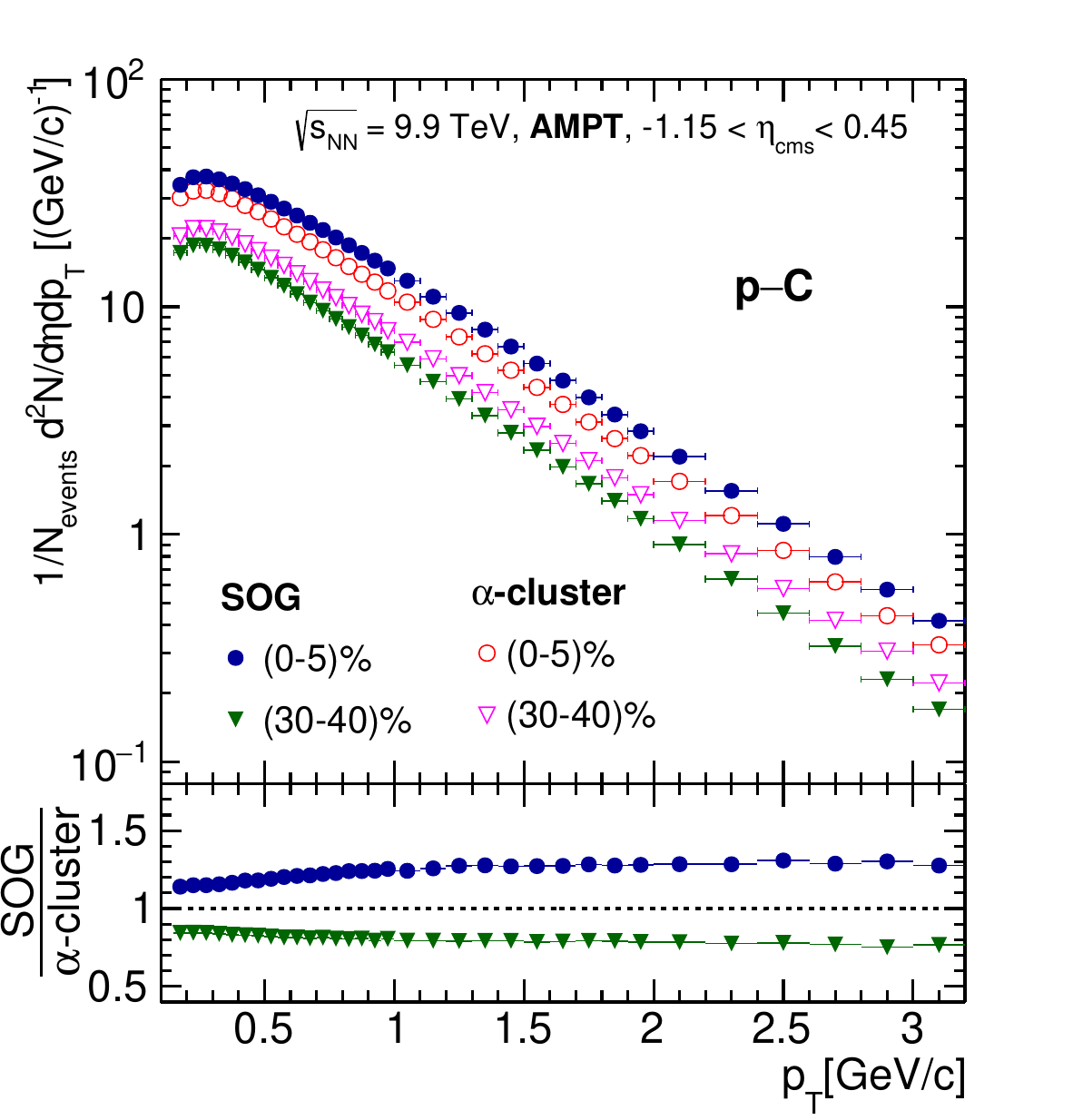}
\caption{(Color online) Upper panels show $p_{\rm T}$-spectra of all charged particles for different nuclear density profiles in p--O (left) and p--C (right) collisions at $\sqrt{s_{\rm NN}}$ = 9.9 TeV for (0--5)\% and (30--40)\%  centrality classes. The ratios of  $p_{\rm T}$ spectra of SOG to $\alpha$--cluster density profile for respective centrality classes are shown in the lower panel.}
\label{fig:pTpOpC}
\end{figure*}
In this section, we begin with a comparison of transverse momentum ($p_{\rm T}$) and pseudorapidity ($\eta$) spectra for the final-state charged particles followed by a discussion on the participant eccentricity ($\epsilon_2$) and triangularity ($\epsilon_3$) in p--C and p--O collisions at $\sqrt{s_{\rm NN}}$ = 9.9 TeV using AMPT. Then, we discuss the centrality dependence of elliptic flow ($v_{2}$) and triangular flow ($v_{3}$), and $v_{3}/v_{2}$ as a function of collision centrality. Finally, the results for $v_{2}/\epsilon_2$ and $v_{3}/\epsilon_3$ as a function of collision centrality have been reported. All these results include a comparison between the SOG and $\alpha$--cluster type density profiles for the $^{12}$C and $^{16}$O nucleus.

As discussed earlier, we define the collision centrality by the geometrical slicing of the impact parameter ($b$) distributions for p--C and p--O collisions at $\sqrt{s_{\rm NN}}$ = 9.9 TeV. Table~\ref{tab:glaubpO} and ~\ref{tab:glaubpC}  show the impact parameter values against each centrality class, along with the mean number of binary collisions ($\langle N_{\rm coll}\rangle$) and average number of participants ($\langle N_{\rm part}\rangle$) for both SOG and $\alpha$--cluster type nuclear density profiles in p--O and p--C collisions at $\sqrt{s_{\rm NN}}$ = 9.9 TeV, respectively. These values are obtained from the Glauber model simulation~\cite{Loizides:2017ack}.

Since the collision systems are asymmetric, the center-of-mass frame of p--O or p--C collisions do not coincide with the laboratory frame. Thus, the shift in rapidity of the center-of-mass frame ($y_{\rm cms}$) from the lab frame ($y_{\rm lab}$) needs to be included. For asymmetric collisions, the shift in rapidity ($\Delta y$) can be expressed in terms of the atomic number ($Z$) and mass number ($A$) of the colliding nuclei as,
\begin{equation}
    \Delta y = |y_{\rm lab}-y_{\rm cms}| = \frac{1}{2} \ln{[Z_{1}A_{2}/Z_{2}A_{1}]}.
\end{equation}
The subscripts `$1$' and `$2$' denote the projectile and target nuclei, respectively. For both p--O and p--C collisions, the rapidity shift is estimated to be $\Delta \rm y$ = 0.346 units in the direction of the proton beam. As a result, a detector coverage of $|\eta_{\rm lab}|<0.8$ would imply the nucleon-nucleon c.m.s to be roughly $-1.15 < \eta_{\rm cms} < 0.45$. Here onwards, $\eta_{\rm lab}$ would be denoted as $\eta$ for simplicity.


\subsection{Transverse momentum (\texorpdfstring{$p_{\rm T}$}{}) and pseudorapidity (\texorpdfstring{$\eta$}{}) distributions} 
\begin{figure}[ht!]
\includegraphics[scale=0.35]{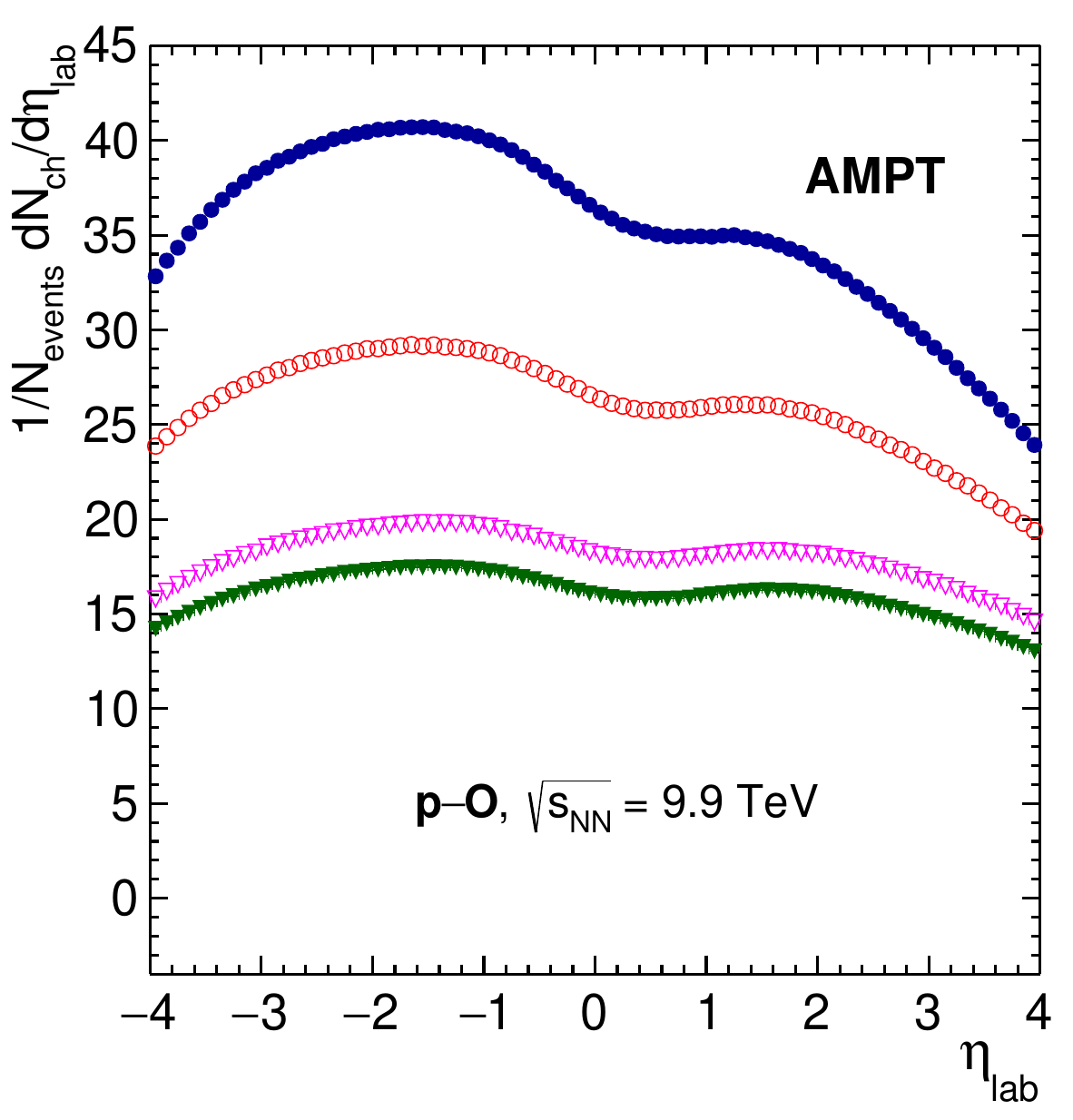}
\includegraphics[scale=0.35]{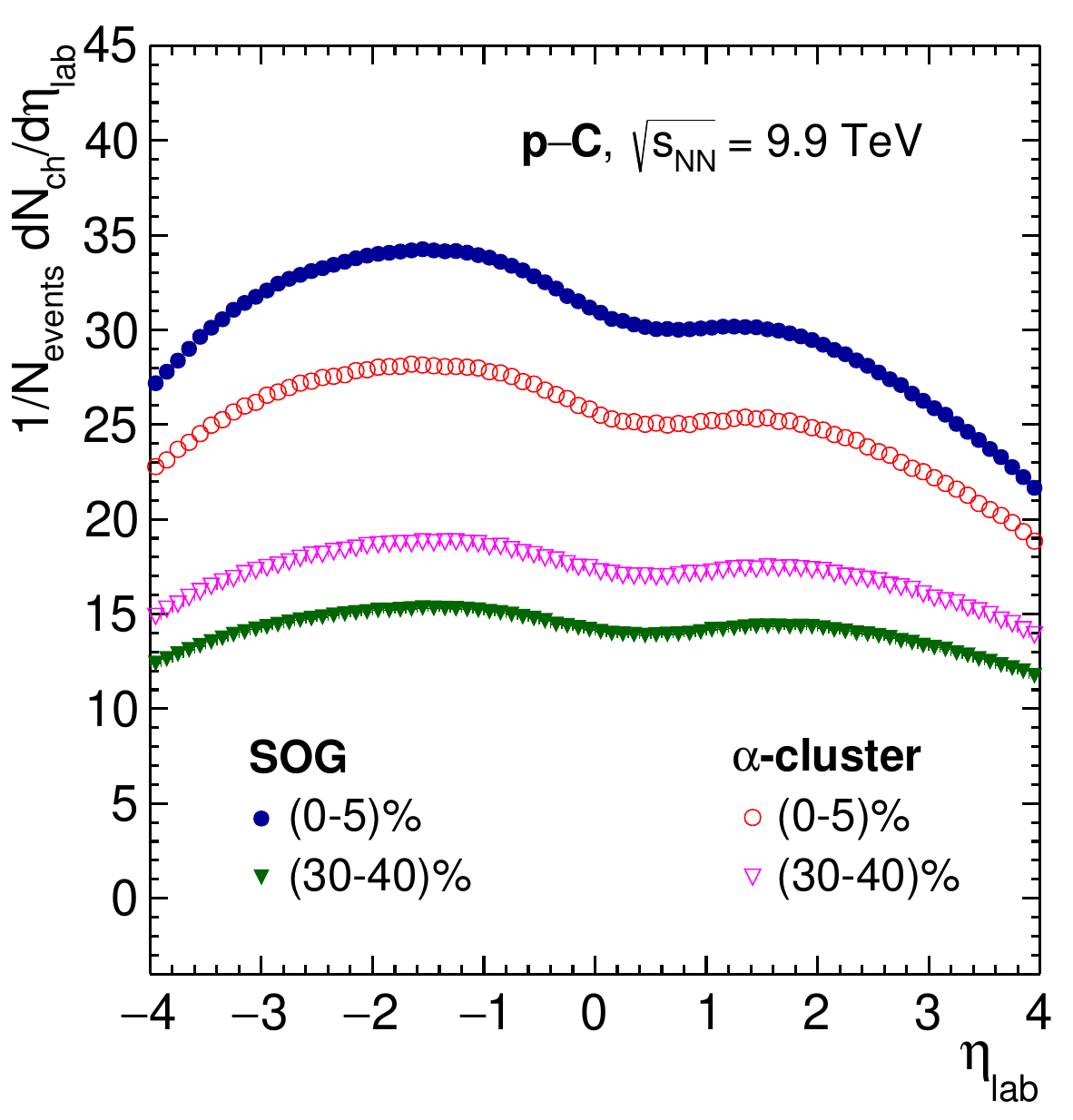}
\caption{(Color online) $\eta$-spectra of charged particles ($p_{\rm T} > 0.15$ GeV/$c$) for different nuclear density profiles in p--O (upper) and p--C (lower) collisions at $\sqrt{s_{\rm NN}}$ = 9.9 TeV  for (0--5)\% and (30--40)\%  centrality classes.}
\label{fig:EtapOpC}
\end{figure}

Figure~\ref{fig:pTpOpC} shows the event normalized $p_{\rm T}$-spectra for unidentified charged hadrons in $|\eta_{\rm lab}| < 0.8$ for central (0--5)\% and mid-central (30--40)\% p--O (left), p--C (right) collisions at $\sqrt{s_{\rm NN}}=9.9$ TeV using AMPT. The $p_{\rm T}$ spectra include the cases for the two nuclear density profiles considered in this study, \textit{i.e.}, the SOG and the $\alpha$--cluster. For the most central p--O and p--C collisions, the charged particle yield is higher for the case of SOG as compared to the $\alpha$--cluster density profile. This enhancement of yield for the SOG density profile for the most central collisions is larger in p--O collisions compared to the same centrality in p--C collisions. In contrast, an opposite trend is observed for the mid-central collisions, where the charged particle yield is enhanced for $\alpha$--cluster density profile compared to SOG. 

On the bottom panels of Fig.~\ref{fig:pTpOpC}, the ratio of $p_{\rm T}$ spectra of SOG to $\alpha$--cluster type density profile has been shown for both collision systems. The ratios show a hardening of $p_{\rm T}$ spectra in the most central collisions for the SOG density profile compared to the $\alpha$--cluster case. On the other hand, for the mid-central collisions, the SOG type density profile produces a slightly softer $p_{\rm T}$-spectra than $\alpha$--cluster density profile. These effects of hardening in the central and softening in the peripheral collisions for the SOG density profile than the $\alpha$--cluster density profile are significantly prominent in p--O than in p--C collisions~\cite{ALICE:2014nqx, ALICE:2012mj}. In hydrodynamics, the hardening of $p_{\rm T}$ spectra is attributed to the collective transverse flow generated due to hydrodynamic expansion~\cite{Bozek:2011if}. As transport models like AMPT incorporate many of the experimentally measured collective features~\cite{Das:2022lqh, Altmann:2024icx, Mallick:2020ium, Mallick:2021hcs, Behera:2021zhi, Mallick:2022alr, Prasad:2022zbr, Behera:2023nwj, Mallick:2023vgi}, a hardening of $p_{\rm T}$ spectra may hint to large values of mean transverse radial flow velocity ($\langle\beta_{\rm T}\rangle$) of the final-state hadrons.

Figure~\ref{fig:EtapOpC} represents the $\eta$-spectra of all charged hadrons with $p_{\rm T} > 0.15$ GeV/$c$ for (0--5)\% and (30--40)\% centrality classes in p--O (upper) and p--C (lower) collisions at $\sqrt{s_{\rm NN}}=9.9$ TeV for SOG and $\alpha$--clustered density profiles. One can observe a similarity in terms of collision centrality and nuclear density profile dependence of charged particle yield between Fig.~\ref{fig:pTpOpC} and \ref{fig:EtapOpC}. The yield in the most central p--O and p--C collisions is higher for the SOG nuclear density profile as compared to the clustered geometry of the nucleons. However, in the mid-central collisions, the yield is higher for $\alpha$--clustered nuclear density profile as compared to the SOG case. 

The reason for the increased yield for asymmetric central collisions with the projectiles having SOG nuclear density profile can be attributed to the varying distribution of the nucleons inside the colliding nuclei. A nucleus with SOG distribution can be pictured as having a gradually decreasing matter density with an increase in distance from the center of the nuclei. However, for the $\alpha$-clustered nuclei, the nucleons stay clustered, leaving physical spaces in between as represented in Fig.~\ref{fig:ALPHApOpC}. Thus, the impact of p--A collisions on the final state can depend on the 3-D orientation of the colliding nucleus `A', which the colliding proton encounters. In central p--A collisions, the proton encounters a denser nuclear medium for the SOG case as compared to an $\alpha$-clustered nuclear profile, leading to an enhanced yield in the former than in the latter. But for mid-central p--A collisions, the yield from the proton colliding with the nuclei having clustered structure nucleus would be greater than that with a nucleus having SOG nuclear profile because the nuclear density is maximum at a certain radius from the centre in the former.

The double-peaks appearing in the pseudorapidity distribution seem to be symmetric around $\eta = 0$ towards peripheral p--O and p--C collisions, which is similar to pp collisions, as expected, possibly due to their similarity in geometry~\cite{Tao:2023kcu}. In other words, due to the smaller number of participants in the mid-central and peripheral p--O and p--C collisions, the scenario becomes equivalent to pp collisions with $\langle N_{\rm part} \rangle \simeq 2-4$. However, as we move towards the central collisions in asymmetric collision systems, the pseudorapidity distribution becomes progressively more asymmetric around $\eta = 0$, such that fewer particles are produced in the direction of the proton beam while there is a predominant emission from the participant nucleons of the heavier nucleus. This is because, unlike A--A nuclear collisions, the scenario in p--A collisions is that of a single nucleon probing the nucleons of a target nucleus in a narrow cylinder~\cite{ALICE:2022imr}. In a central p--A collision, the projectile proton interacts with a denser volume of the target nucleus than in the case of a peripheral p--A collision. Consequently, the yield in the nucleus-beam-going direction will increase from peripheral to central collisions. As a result, one can see in Fig.~\ref{fig:EtapOpC} that the multitude of particles produced is higher in the oxygen- and carbon-going directions than in the proton-going direction for the most central p--O and p--C collisions, respectively. In addition, one observes a slightly higher asymmetry in the double-peak structure of pseudorapidity distribution in p--O collisions as compared to p--C collisions, presumably due to a denser $^{16}$O nuclei. The effects are further enhanced for an SOG system as compared to $\alpha$--cluster nuclear density profile, indicating that the core part of an SOG is slightly denser as compared to an $\alpha$--cluster nuclear geometry. 

\subsection{Eccentricity and triangularity}

\begin{figure*}[ht!]
\includegraphics[scale=0.42]{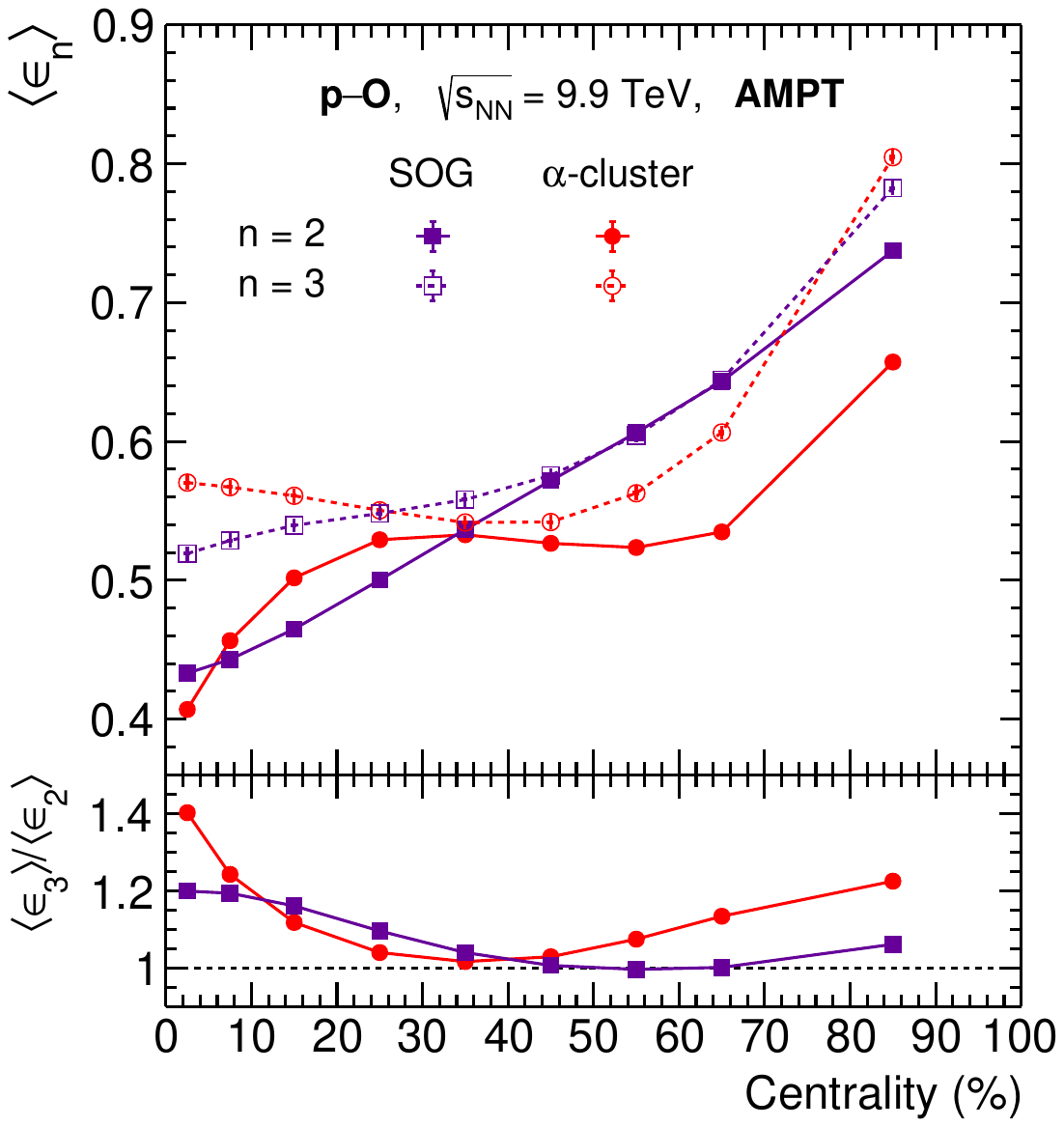}
\includegraphics[scale=0.42]{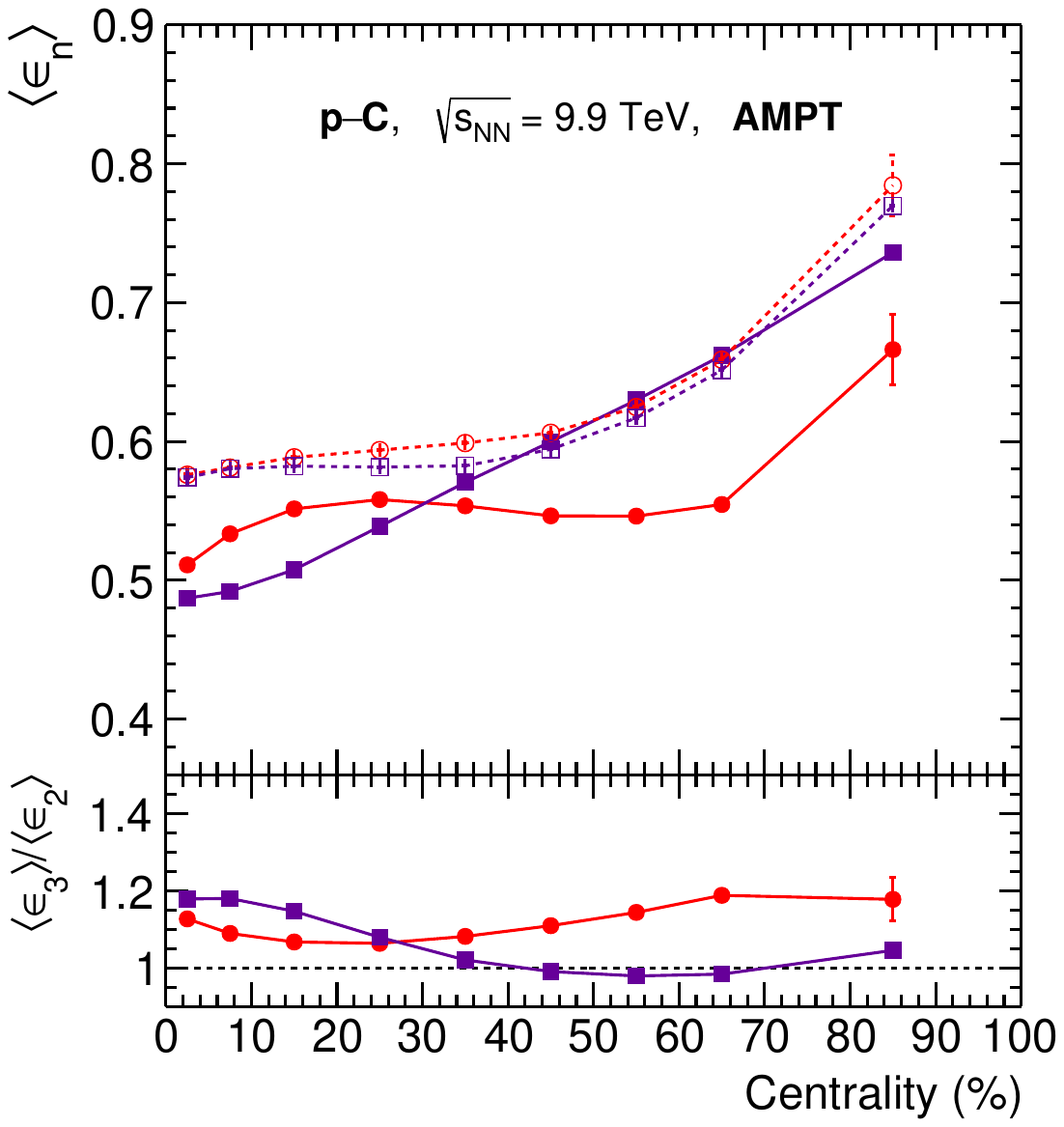}
\caption{(Color online) Upper panels show the centrality dependence of average eccentricity ($\langle\epsilon_{2}\rangle$) and triangularity ($\langle\epsilon_{3}\rangle$) for p--O (left) and p--C (right) collisions at $\sqrt{s_{\rm NN}}$ = 9.9 TeV using AMPT for SOG and $\alpha$--cluster nuclear density profiles. The ratio $\langle\epsilon_{3}\rangle$/$\langle\epsilon_{2}\rangle$ is plotted as a function of centrality in the corresponding lower panels.}
\label{fig:en}
\end{figure*}

In non-central nuclear collisions, the almond-shaped geometry of the nuclear overlap region creates initial spatial anisotropy, which induces asymmetric pressure gradients in the transverse plane. This spatial anisotropy in the initial state gets transformed into the azimuthal momentum anisotropy in the final state and appears as the collective expansion of the medium. This initial spatial anisotropy can be quantified in terms of eccentricity ($\epsilon_{2}$) and triangularity ($\epsilon_{3}$) of the participant nucleons. Eccentricity quantifies the extent to which the transverse overlap region is elliptical in shape, and triangularity tells how triangular the transverse participant plane is. While eccentricity has a major contribution from the asymmetry in the collision geometry, triangularity can develop due to the event-by-event density fluctuations in the collision overlap region. In experiments, the measurement of $\epsilon_{2}$ and $\epsilon_{3}$ is nontrivial. However, Monte Carlo models like AMPT can provide the coordinates of the participant nucleons at the instance of collision. Thus, one can quantify the spatial anisotropy, $\epsilon_{\rm n}$, for the $n^{\rm th}$ order harmonics, using the following expression~\cite{Petersen:2010cw},
\begin{equation}
\epsilon_{n} = \frac{\sqrt{ \langle r^{n} \cos({n \phi_{\rm part}}) \rangle^{2} + \langle r^{n} \sin({n \phi_{\rm part}}) \rangle^{2} }}{\langle r^{n}\rangle}.
\label{eq:en}
\end{equation}
Here, $r$ and $\phi_{\rm part}$ are the polar coordinates of the participant nucleons in the transverse plane. $\epsilon_{2}$ and $\epsilon_{3}$ correspond to the second and third order spatial anisotropy, respectively. In Eq.~\eqref{eq:en}, $\langle \dots \rangle$ denotes the average taken over all the participating nucleons in an event. Although the initial state in A--A collisions can be characterized by the overall geometrical shape of the interaction region, the scenario becomes different in asymmetric p--A collisions. In the latter, the number of participants is comparatively less than nucleus-nucleus collisions, and the system's geometry grows sensitive to the proton's structure and size.
This means that the source of both eccentricity and triangularity in proton-induced interactions are fluctuation-dominated, and the size of the system is governed by the incoming proton~\cite{Bzdak:2013zma}.


The upper panels in Fig.~\ref{fig:en} show the event-averaged eccentricity ($\langle\epsilon_{2}\rangle$) and triangularity ($\langle\epsilon_{3}\rangle$) as a function of centrality for p--O (left) and p--C (right) collisions at $\sqrt{s_{\rm NN}}$ = 9.9 TeV while the ratio $\langle\epsilon_{3}\rangle/\langle\epsilon_{2}\rangle$ is plotted in the corresponding lower panels. For the case of SOG density profile in both p--O and p--C collision systems, $\langle\epsilon_{2}\rangle$ increases linearly with an increase in centrality. This trend is also quite similar to that of $\langle\epsilon_{2}\rangle$ for the Woods-Saxon density profile in O--O collisions obtained using AMPT~\cite{Behera:2023nwj}. In A--A collisions, the reason for the increasing trend of $\langle\epsilon_{2}\rangle$ with increasing centrality percentile is a growing elliptic geometry of the collision overlap region. However, in p--A collisions, eccentricity can increase from central to peripheral collisions due to the contributions from increasing fluctuations or decreasing nuclear participants with increasing impact parameter.

For the case of $\alpha$--cluster type density profile in both p--O and p--C collision systems, the behavior of $\langle\epsilon_{2}\rangle$ with changing collision centrality becomes much more interesting. Here, $\langle\epsilon_{2}\rangle$ has a non-linearly rising curvy trend till (20--30)\% centrality, which then remains almost unchanged with increasing centrality up to (60--70)\%, and finally shows a sharp rise for the (70--100)\% centrality bin. This behavior of $\langle\epsilon_{2}\rangle$ with an increase in collision centrality is similar for both p--O and p--C collisions for the $\alpha$--cluster case. In addition, a similar trend of $\langle\epsilon_{2}\rangle$ with centrality is also observed for O--O collisions with an $\alpha$--cluster nuclear density profile using AMPT~\cite{Behera:2023nwj}. This observation of different trends of $\langle\epsilon_{2}\rangle$ as a function of collision centrality for the nuclei having SOG or $\alpha$--cluster type density profiles, indicates that not only the colliding nuclear species but also its nuclear density profile, \textit{i.e.,} the way nucleons are distributed inside the nucleus plays a major role in shaping the initial spatial anisotropy of the collision overlap region.




The variation of $\langle\epsilon_{3}\rangle$ as a function of collision centrality for p--O and p--C collisions is also shown in Fig.~\ref{fig:en}. In p--C collisions, $\langle\epsilon_{3}\rangle$ remains almost flat up to the mid-central collisions and then shows a linear increasing trend towards the peripheral collisions. This trend is almost indistinguishable for both SOG and $\alpha$-cluster type density profiles in p--C collisions. In contrast, in p--O collisions, $\langle\epsilon_{3}\rangle$ for $\alpha$--cluster density profile dominates over that of SOG density profile in the most central case, drops in the mid-central collisions followed by a rise in the peripheral cases. As this pattern of $\langle\epsilon_{3}\rangle$ for $\alpha$--cluster density profile in p--O collisions is similar to that of $\langle\epsilon_{3}\rangle$ in Ref.~\cite{Behera:2023nwj} for O--O collisions, this trend could be attributed specifically to the presence of an $\alpha$--clustered $^{16}\rm O$ nucleus. 

The bottom panels of Fig.~\ref{fig:en} show $\langle\epsilon_{3}\rangle/\langle\epsilon_{2}\rangle$ for a given density profile. For the case of SOG density profile in both p--O and p--C collision systems, $\langle\epsilon_{3}\rangle/\langle\epsilon_{2}\rangle$ shows a very similar trend, where it is greater than unity for the most central collisions, which then gradually decreases to approach unity ($\langle\epsilon_{3}\rangle \simeq \langle\epsilon_{2}\rangle$) for mid-central to peripheral collisions. This trend indicates that the spatial anisotropy in the most central p--A collisions is mostly dominated by event-by-event fluctuations as compared to the shape of the nuclear overlap region since $\langle\epsilon_{3}\rangle/\langle\epsilon_{2}\rangle > 1.0$. From mid-central to peripheral collisions, the greater rise in $\langle\epsilon_{2}\rangle$ brings down the ratio to unity, which means, both $\langle\epsilon_{2}\rangle$ and $\langle\epsilon_{3}\rangle$ contribute equally to the initial spatial anisotropy in peripheral collisions. 

For the $\alpha$--cluster case in p--O collisions, $\langle\epsilon_{3}\rangle/\langle\epsilon_{2}\rangle$ shows a sharp rise from mid-central to central collision which even exceeds the value of $\langle\epsilon_{3}\rangle/\langle\epsilon_{2}\rangle$ for the SOG type profile in the most central collisions. This feature is once again consistent with the results for O--O collisions for $\alpha$--cluster density profile using AMPT~\cite{Behera:2023nwj}. The dominance of $\langle\epsilon_{3}\rangle/\langle\epsilon_{2}\rangle$ for the most central p--O collisions with the $\alpha$--clustered $^{16}$O nuclear density profile, is absent in the corresponding p--C collisions. This behavior may be thus attributed to the presence of an extra $\alpha$--cluster in $^{16}$O compared to $^{12}$C, which leads to some additional density fluctuations in the most central p--O collisions resulting in a higher value of $\langle\epsilon_{3}\rangle$. This could be understood as follows. The tetrahedral nuclear structure of $^{16}\rm O$ has four triangular surfaces in comparison to the $^{12}\rm C$ nucleus, which has only one triangular surface. Thus, the presence of an additional alpha-particle in the clustered $^{16}\rm O$ nuclei can give rise to an increase in triangularity fluctuations, which increases $\langle\epsilon_{3}\rangle$ in the most central p--O collisions. This is similar to the case of $^{238}\rm U$, where its nuclear deformation leads to a rise in ellipticity fluctuations in the most central U--U collisions~\cite{Giacalone:2021udy}. Therefore, the observed sharp rising trend of $\langle\epsilon_{3}\rangle/\langle\epsilon_{2}\rangle$ towards the most central p--O and O--O collisions at the LHC energy can be strongly attributed to the presence of tetrahedral $\alpha$--cluster geometry in $^{16}$O.

\begin{figure}[ht!]
\includegraphics[scale=0.43]{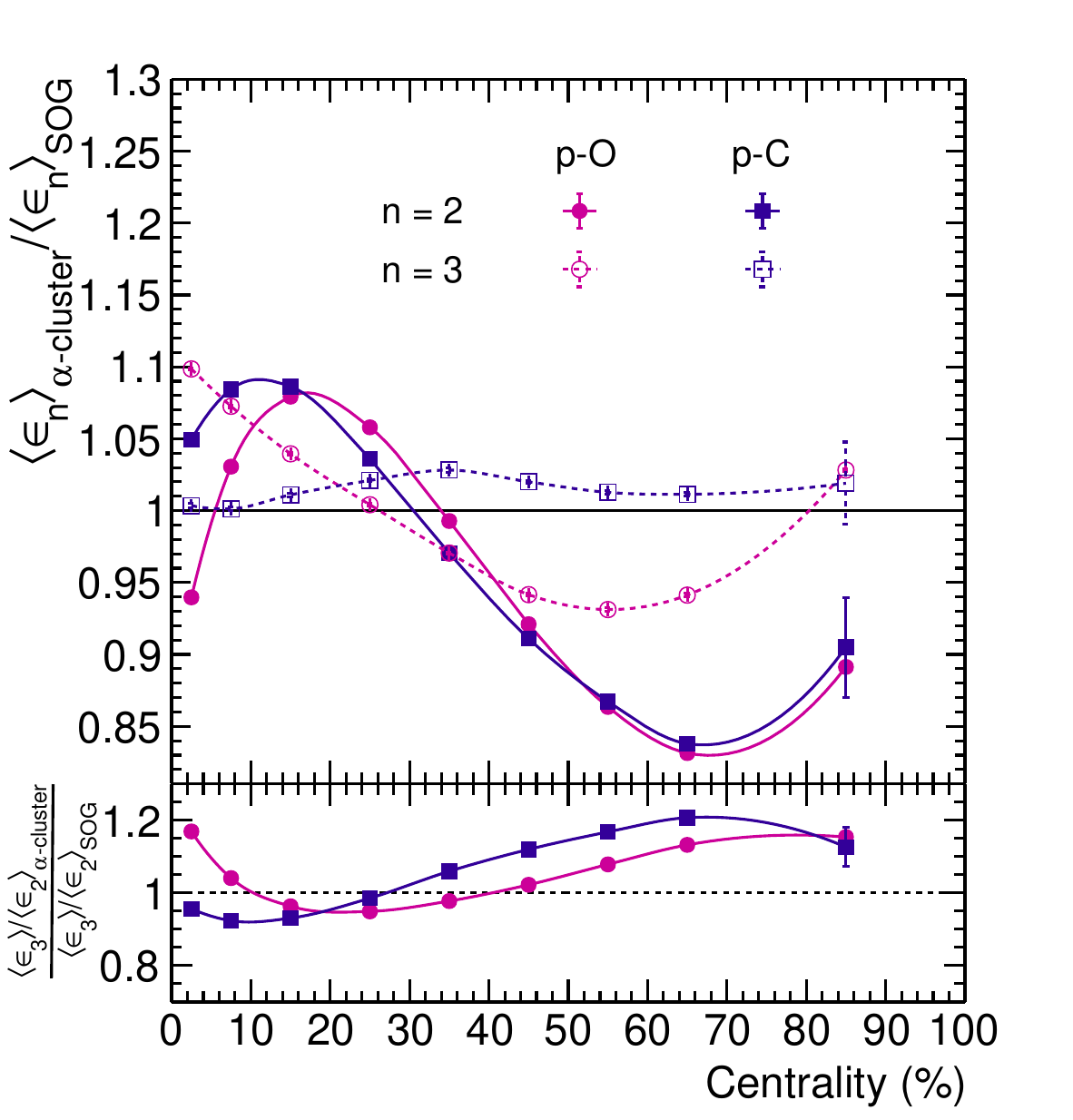}
\caption{(Color online) Comparison of $\langle\epsilon_{2}\rangle$, $\langle\epsilon_{3}\rangle$ (upper panel) and $\langle\epsilon_{3}\rangle$/$\langle\epsilon_{2}\rangle$ (lower panel) between $\alpha$-cluster and SOG density profiles depicted through ratios, for p--O and p--C collisions at $\sqrt{s_{\rm NN}}$ = 9.9 TeV using AMPT.}
\label{fig:enalphasog}
\end{figure}

In order to have a clearer comparison between the effects due to clustered and unclustered nuclear density profiles, a ratio between the two density profiles for the studied observables is meaningful. The upper panel of Fig.~\ref{fig:enalphasog} shows the ratio of average eccentricity ($\langle\epsilon_2\rangle$) and triangularity ($\langle\epsilon_3\rangle$) of $\alpha$--clustered nuclear profile to that of SOG type density profile as a function of centrality in p--O and p--C collisions at $\sqrt{s_{\rm NN}}$~=~9.9~TeV using AMPT. For the case of eccentricity, both p--O and p--C collisions show an enhancement in the ratio value from most central to (10--20)\%, which then monotonously decreases towards peripheral collisions, with a small rise for the (70--100)\% centrality bin. For triangularity, the ratio between $\alpha$--cluster and SOG profile remains close to unity for p--C collisions; however, the same ratio of triangularity shows a sharp linear rise from mid-central to most central bin due to the sharp rise in $\langle\epsilon_3\rangle$ for $\alpha$--clustered case in p--O collisions as discussed in Fig.~\ref{fig:en}. The lower panel of Fig.~\ref{fig:enalphasog} compares the ratio of $\langle\epsilon_3\rangle/\langle\epsilon_2\rangle$ from $\alpha$-cluster with SOG nuclear profile as a function of collision centrality in p--O and p--C collisions. Again, the sharp rise in this ratio in central p--O collisions is due to the dominance of the numerator, \textit{i.e.,} $\langle\epsilon_3\rangle/\langle\epsilon_2\rangle$ for the $\alpha$--clustered nuclear density profile. On the other hand, no such sharp rise effect is seen for the p--C collisions. 
 
\subsection{Elliptic flow and triangular flow}


\begin{figure}[ht!]
\includegraphics[scale=0.5]{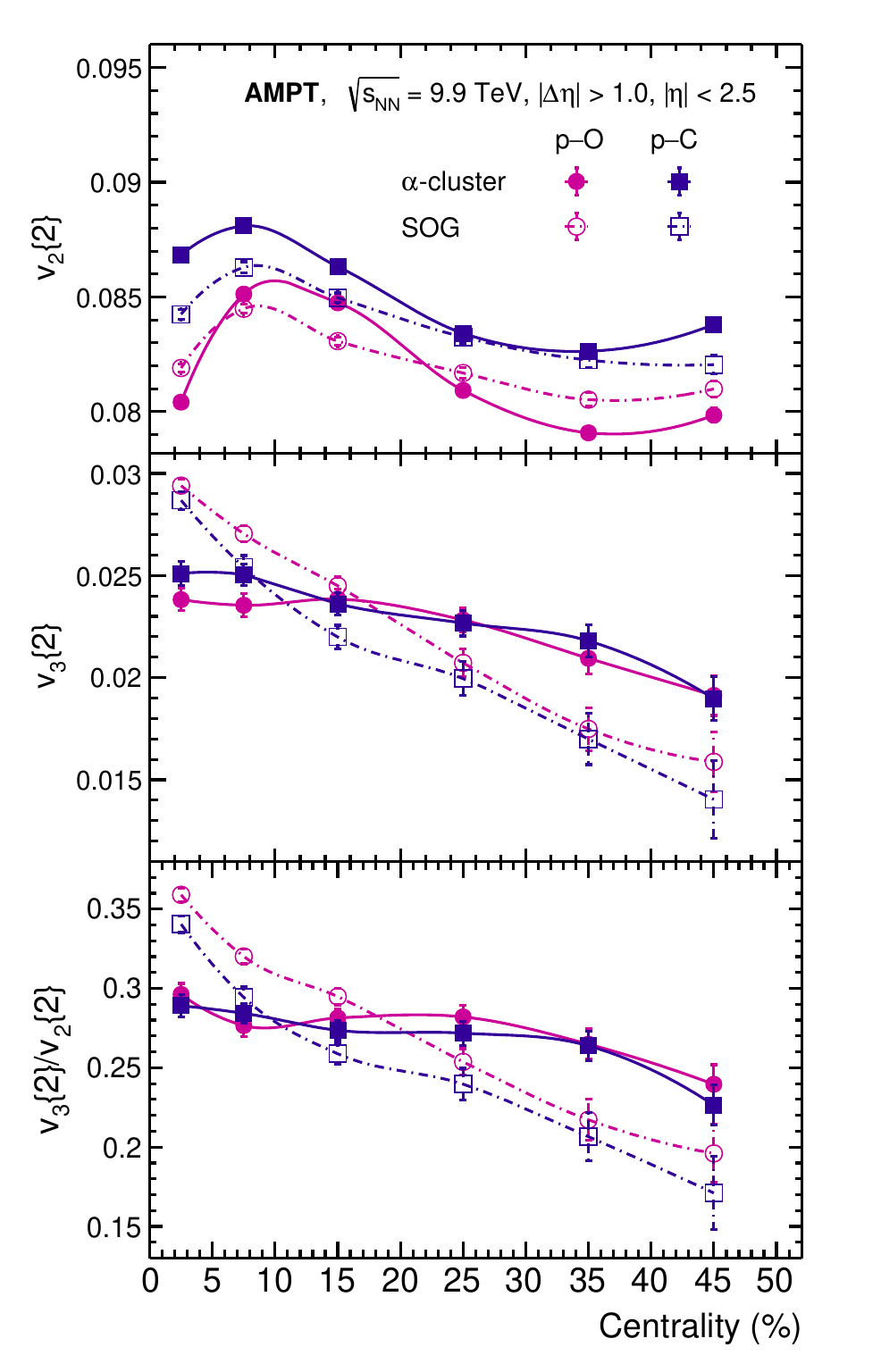}
\caption{(Color online) Centrality dependence of elliptic flow ($v_{2}\{2\}$) (upper), triangular flow ($v_{3}\{2\}$) (middle) and their ratio $v_{3}\{2\}$/$v_{2}\{2\}$ (lower) for p--O and p--C collisions at $\sqrt{s_{\rm NN}}$~=~9.9 TeV for SOG and $\alpha$--cluster nuclear density profiles using AMPT model.}
\label{fig:v2v3v3byv2}
\end{figure}

\begin{figure}[ht!]
\includegraphics[scale=0.43]{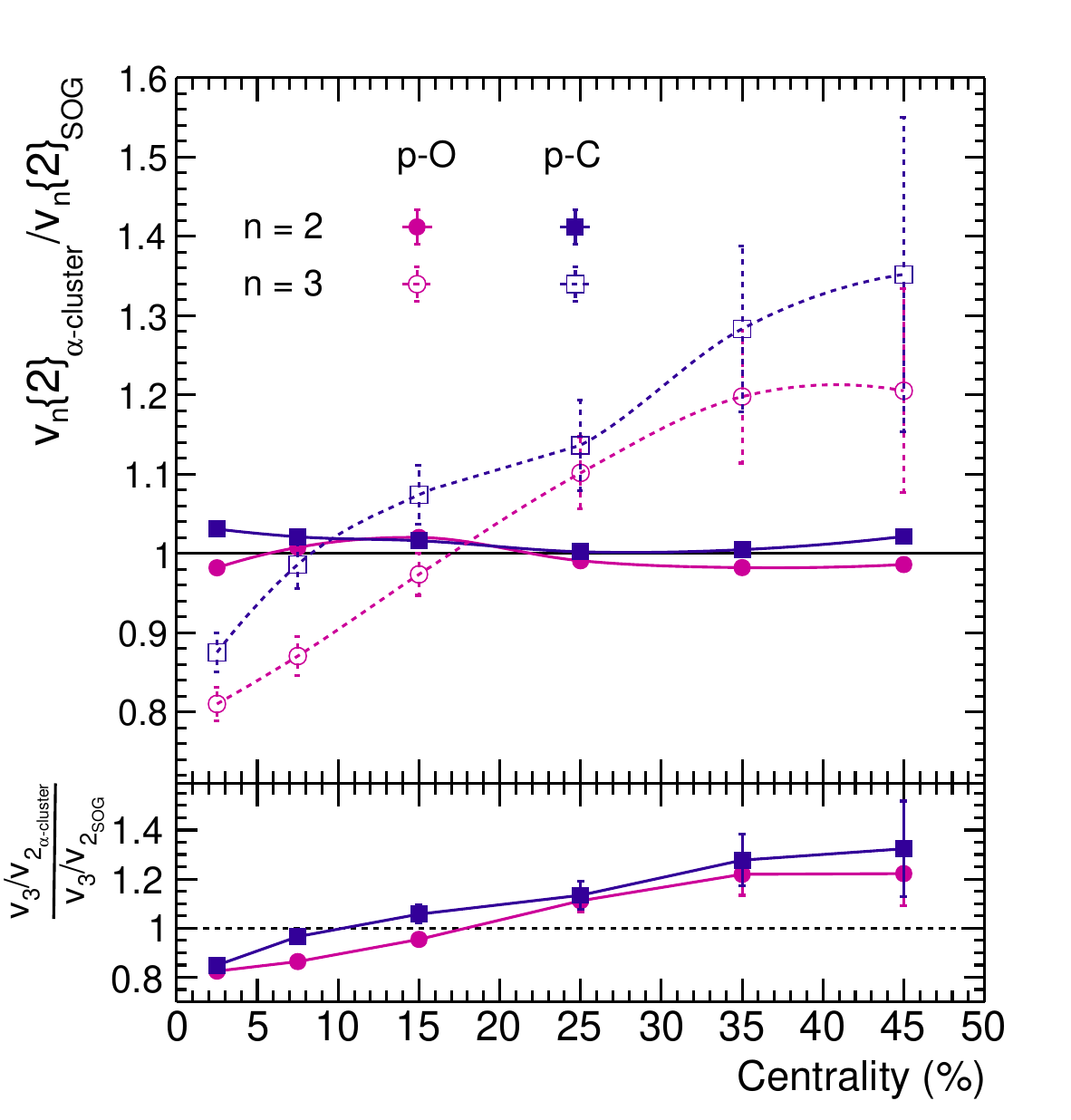}
\caption{(Color online) Comparison of $v_{2}\{2\}$, $v_{3}\{2\}$ and $v_{3}\{2\}$/$v_{2}\{2\}$ between $\alpha$-cluster and SOG density profiles depicted through ratios, for p--O and p--C collisions at $\sqrt{s_{\rm NN}}$~=~9.9 TeV using AMPT.}
\label{fig:vnbyvnalphasog}
\end{figure}

\begin{figure*}[ht!]
\includegraphics[scale=0.35]{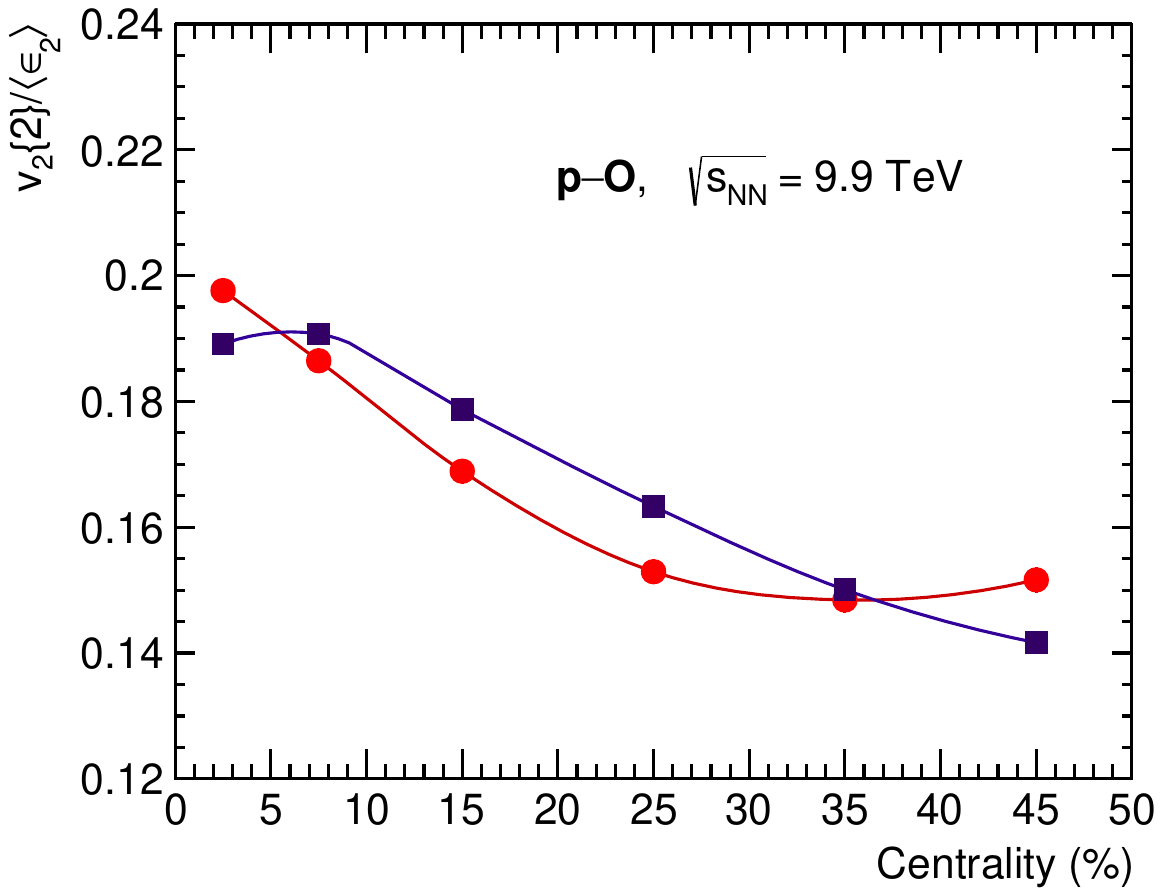}
\includegraphics[scale=0.35]{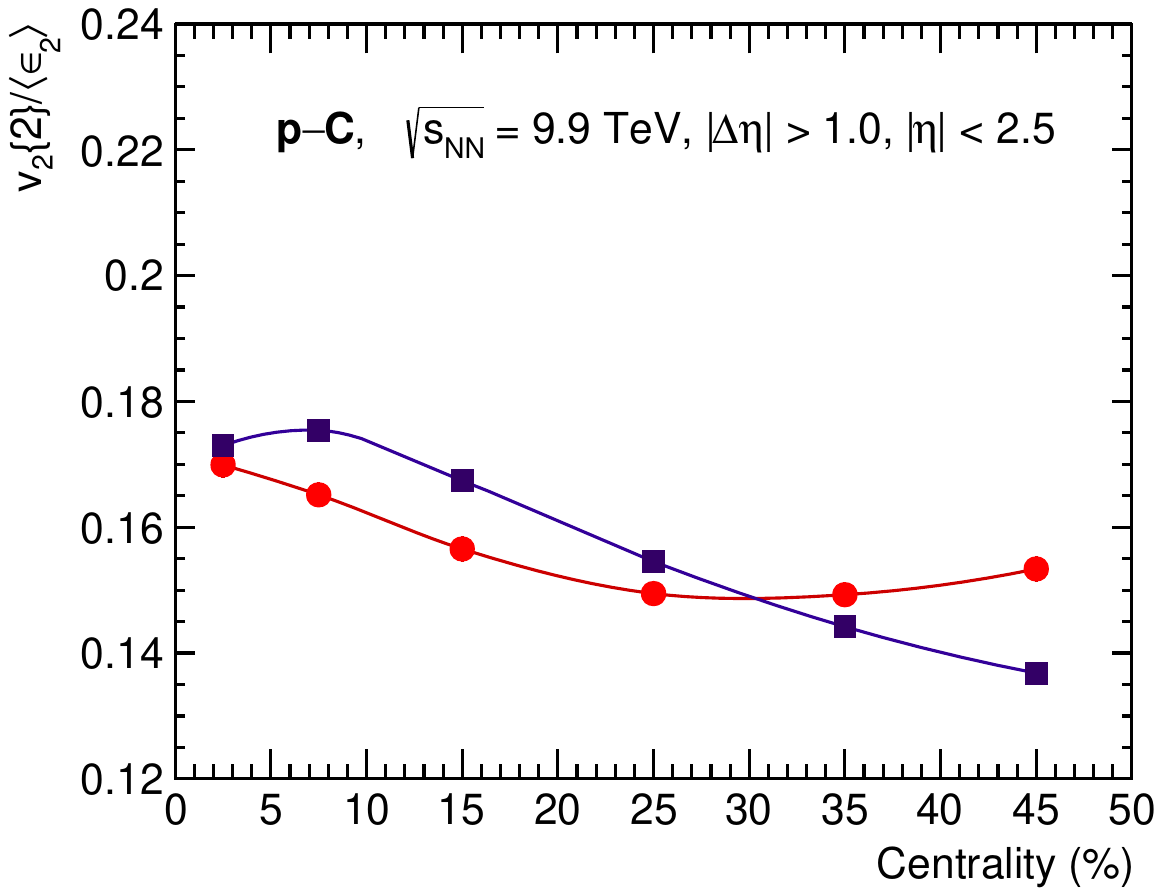}
\includegraphics[scale=0.35]{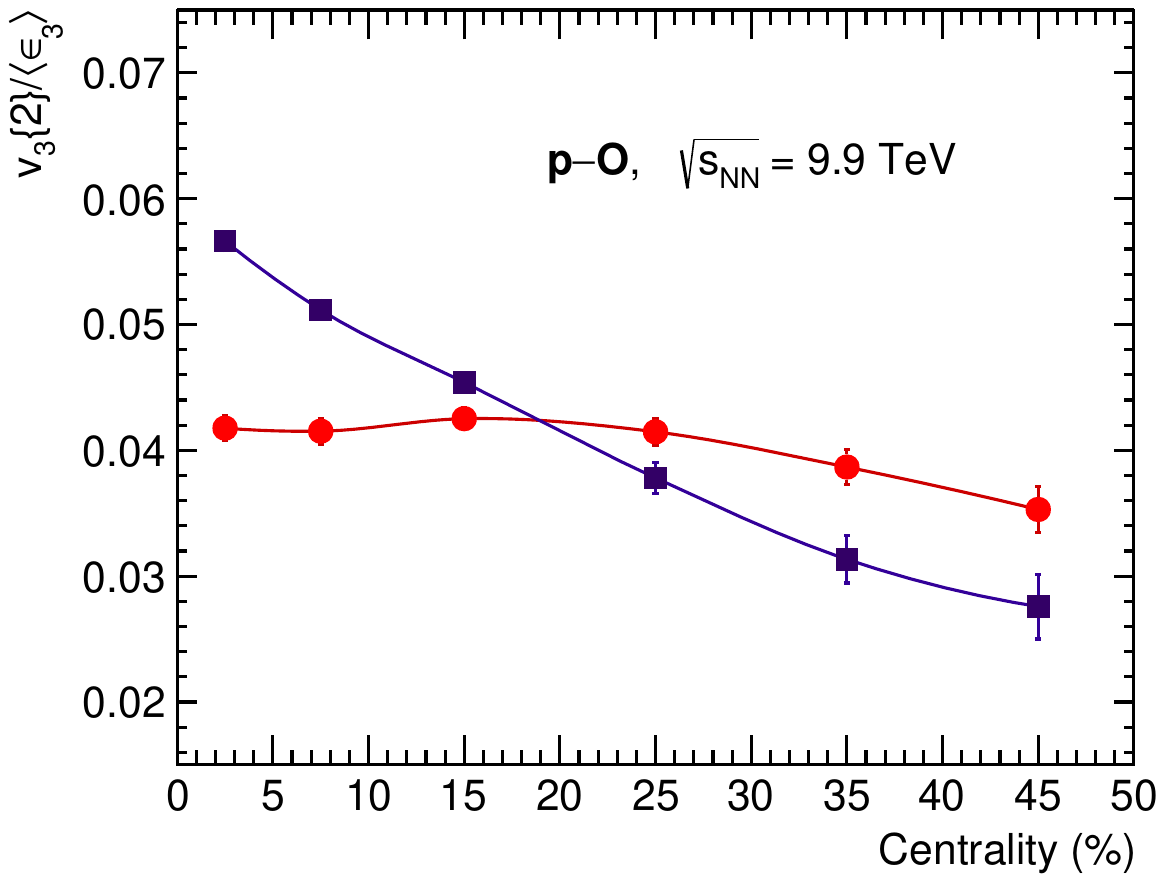}
\includegraphics[scale=0.35]{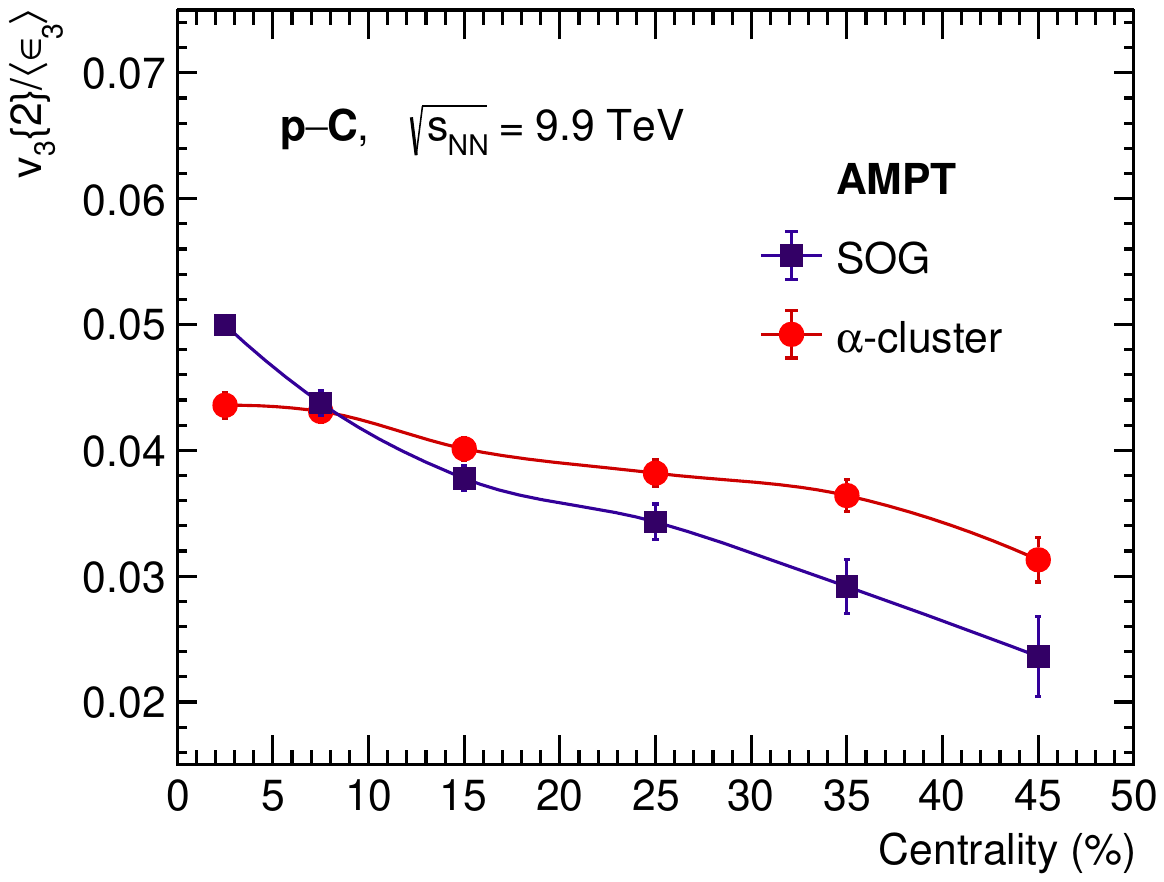}
\caption{(Color online) Ratio $v_{2}\{2\}/\langle\epsilon_2\rangle$ (upper) and $v_{3}\{2\}/\langle\epsilon_3\rangle$ (lower) as a function of centrality in p--O (left) and p--C (right) collisions at $\sqrt{s_{\rm NN}}=9.9$ TeV using AMPT for both SOG and $\alpha$--cluster nuclear density profiles.}
\label{fig:vnen}
\end{figure*}
The flow coefficients are estimated using the two-particle Q-cumulant method with a relative pseudorapidity gap, $|\Delta\eta|>1.0$ between the particle pairs. The details of the Q-cumulant method used in the study can be found in Appendix~\ref{qcumulant}. Figure~\ref{fig:v2v3v3byv2} shows the centrality dependence of elliptic flow ($v_{2}\{2\}$) (upper panel), triangular flow ($v_{3}\{2\}$) (middle panel) and the ratio $v_{3}\{2\}/v_{2}\{2\}$ for all charged hadrons in $|\eta| < 2.5$ for SOG and $\alpha$--cluster nuclear density profiles in p--O and p--C collisions at $\sqrt{s_{\rm NN}}=9.9$~TeV using AMPT. 

For the case of SOG type density profile, a similar trend of centrality dependence of $v_{2}\{2\}$ for both p--O and p--C collisions is observed; however, the elliptic flow estimated from p--C collisions is approximately 2.5\% higher compared to p--O collisions. The value of $v_{2}\{2\}$ is found to rise from the most central to (5--10)\% centrality class, followed by a smooth decrease towards the mid-central and peripheral collisions. Again, one observes a similar centrality dependence curve of $v_{2}\{2\}$ between p--O and p--C collisions with the $\alpha$--cluster nuclear density profile. Overall, p--C collisions result in approximately 8.5\% higher values elliptic flow than p--O collisions for $\alpha$--cluster case. This centrality dependence of elliptic flow coefficients, though weak, is consistent with the experimental observation of elliptic flow in p--Pb~\cite{ALICE:2019zfl} collisions. The presence of $\alpha$--clusters in p--C collisions, increases the value of $v_2\{2\}$ for the most central collisions up to (10--20)\% centrality class. In p--O collisions, the presence of $\alpha$--cluster results in a smaller value of elliptic flow for all centrality classes except (5-10)\% and (10--20)\% cases. This behavior of $v_2\{2\}$ for the central collisions for different density profiles mimics the behavior of initial eccentricity, thus, it can be attributed to the initial state effect. For the mid-central to peripheral collisions, elliptic flow shows a decreasing trend irrespective of the increasing eccentricity value. This may be attributed to the decreasing number of participants and a smaller time of evolution in peripheral collisions, which suppresses the development of $v_2\{2\}$. Thus, the centrality dependence of $\langle\epsilon_2\rangle$ is not well reflected in the corresponding values of $v_{2}\{2\}$. Overall, the higher values of  $v_{2}\{2\}$ in p--C collisions compared to p--O collisions can be associated with the initial state effect due to the higher value of $\langle\epsilon_2\rangle$ for the p--C collisions as seen in Fig.~\ref{fig:en}.





The triangular flow ($v_{3}\{2\}$) as shown in the middle row of Fig.~\ref{fig:v2v3v3byv2}, for both $\alpha$--cluster and SOG nuclear density profiles, is found to decrease almost linearly from central to peripheral p--O and p--C collisions; however, in the presence of $\alpha$--clusters, the curve is less steeper. This results in a higher value of $v_{3}\{2\}$ for the mid-central to peripheral p--O and p--C collisions involving $\alpha$--clusters. One also observes that while $\langle\epsilon_3\rangle$ shows an increasing trend towards the peripheral collisions, $v_{3}\{2\}$ shows a decreasing trend, which is expected in small collision systems like p--O and p--C collisions. For p--C collisions, 
$\langle\epsilon_3\rangle$ for SOG and $\alpha$--cluster case are indistinguishable, yet, the resulting $v_{3}\{2\}$ has sizable quantitative difference. Thus, the evolution from $\langle\epsilon_3\rangle$ to $v_{3}\{2\}$ in small collision systems may get influenced from final state effects unlike $v_{2}\{2\}$, which is mostly driven by initial state effects as discussed above.

In the lower panel of Fig.~\ref{fig:v2v3v3byv2}, one finds that the ratio $v_{3}\{2\}/v_{2}\{2\}$ is strictly less than one, which is consistent with the observations made in Pb--Pb and p-Pb collisions in experiments~\cite{ALICE:2019zfl}. We observe that the centrality dependence of $v_{3}\{2\}$/$v_{2}\{2\}$ qualitatively resembles the centrality dependence of $\langle v_{3}\rangle$, where a gradually decreasing ratio of SOG density profile is cut across in the mid-central classes by an almost flat curve of the $\alpha$--cluster density profile, for both the collision systems. This is because $v_{2}\{2\}$ has a smaller centrality dependence, whereas $v_{3}\{2\}$ is affected significantly by a change in collision centrality, as discussed earlier.


The upper panel of Fig.~\ref{fig:vnbyvnalphasog} shows the centrality dependence of the ratio of $v_{n}\{2\}$ from an $\alpha$--cluster and SOG nuclear density profiles in p--O and p--C collisions at $\sqrt{s_{\rm NN}}=9.9$~TeV using AMPT. Here, we find negligible, less than 5\%, variation in the centrality dependence of $v_{2}\{2\}$ for the $\alpha$--cluster case as compared to the SOG density profile. However, the effect of clustered structure is strongly visible in triangular flow, which amounts to more than 20\% variation. In the lower panel of Fig.~\ref{fig:vnbyvnalphasog}, the ratio of $v_{3}\{2\}/v_{2}\{2\}$ from $\alpha$--clustered structure to SOG density profile is shown as a function of collision centrality in p--O and p--C collisions at $\sqrt{s_{\rm NN}}=9.9$~TeV using AMPT. Here, one finds that the ratio of $v_{3}\{2\}/v_{2}\{2\}$ from $\alpha$--cluster case to SOG density profile varies about 20\%. In addition, the ratio closely resembles the centrality dependence observed in the lower panel of Fig.~\ref{fig:enalphasog}, except for the central cases of p--O collisions.

The ratio $v_n/\epsilon_n$ can be considered as the representation of the medium response to the initial spatial anisotropy.
Figure~\ref{fig:vnen} shows the centrality dependence of the ratio $v_{2}\{2\}/\langle\epsilon_2\rangle$ (upper plots) and $v_{3}\{2\}/\langle\epsilon_3\rangle$ (lower plots) in p--O (left) and p--C (right) collisions at $\sqrt{s_{\rm NN}}=9.9$~TeV using AMPT for both SOG and $\alpha$--cluster nuclear density profiles. Naively, one finds that the centrality dependence of $v_{2}\{2\}/\langle\epsilon_2\rangle$ is nearly proportional to $1/\langle\epsilon_2\rangle$ for both the nuclear density profiles in both p--O and p--C collisions. This is because $v_{2}\{2\}$ has a weak dependence on collision centrality (Fig.~\ref{fig:v2v3v3byv2}), whereas $\langle\epsilon_2\rangle$ shows a relatively stronger centrality dependence, as shown in Fig.~\ref{fig:en}. At the same time, if we observe the centrality dependence of $v_{3}\{2\}/\langle\epsilon_3\rangle$, the SOG nuclear density profiles are found to show a decreasing behavior from central to mid-central collisions, indicating a significant impact from final state medium effects in the evolution of triangularity. However, for the $\alpha$--cluster case, the ratio $v_{3}\{2\}/\langle\epsilon_3\rangle$ is less sensitive to collision centrality in both p--O and p--C collisions where the sensitivity is even less for p--O collisions than that for p--C collisions. These contrasting trends between the two nuclear density profiles mark one of the major observations in this study on p--O and p--C collisions, which signifies the effects due to the presence (or absence) of a clustered nuclear geometry in $^{16}$O and $^{12}$C nuclei.

\section{Summary}
\label{sec4}
For the first time, we report a systematic study of the effects of the nuclear density profiles on the final-state medium anisotropy through p--O and p--C collisions at the LHC. We study the transverse momentum ($p_{\rm T}$) and pseudorapidity ($\eta$) spectra, eccentricity and triangularity, elliptic and triangular flow in p--O and p--C collisions at $\sqrt{s_{\rm NN}}=9.9$~TeV using AMPT with $\alpha$--cluster type geometry and a model-independent Sum of Gaussians (SOG) type nuclear density profiles. The major findings of this study are summarized below:
\begin{itemize}
\item With the study of $p_{\rm T}$ and $\eta$ spectra, the study establishes the differences in the yield for both the nuclear density profiles for different centrality classes. 
\item One finds a significant $\langle \epsilon_n \rangle$ dependence on the collision centrality, where an $\alpha$--cluster nuclear density profile maintains a similar qualitative behavior with an increase in collision centrality in p--O, p--C and O--O collisions. This is one of the important findings of this paper; although it cannot be confronted in experiments, it retains significant importance in understanding the collision overlap region for an $\alpha$--clustered geometry, especially in p--A collisions. 
\item Like in p--Pb collisions, the p--O and p--C collisions with the $\alpha$--cluster nuclear density profile show a weak dependence of triangular flow on collision centrality. In contrast, the triangular flow for the SOG density profile shows a strong centrality dependence for both p--O and p--C collision systems. 

\item Finally, for an $\alpha$--cluster nuclear density profile, $v_{3}\{2\}/\langle\epsilon_3\rangle$ has a small dependence on collision centrality while SOG nuclear profile shows a linear drop with an increase in collision centrality. 


\end{itemize}

With this study using AMPT, we find that the dependence of the nuclear density profile is small in elliptic flow. However, we observe a strong effect of $\alpha$-clustered structure in the final-state triangular flow in both p--O and p--C collisions using AMPT, which is reflected in several observables shown in this study. When confronted with experimental measurements, this study can provide a crucial understanding of the effect of the presence of an $\alpha$--clustered structure of $^{16}$O and $^{12}$C through the p--O and p--C collisions at the LHC energies.


\section*{Acknowledgement}
A.M.K.R. acknowledges the doctoral fellowships from the DST INSPIRE program of the Government of India.
S.P. acknowledges the University Grants Commission (UGC), Government of India. The authors gratefully acknowledge the DAE-DST, Government of India funding under the mega-science project “Indian participation in the ALICE experiment at CERN” bearing Project No. SR/MF/PS-02/2021-IITI(E-37123).

\appendix
\section{}
\label{Appendix-I}

\subsection{A multi-phase transport model}
AMPT is a widely used Monte Carlo transport model constructed to describe the space-time evolution of pp, p+A, and A+A collisions across RHIC and LHC energies~\cite{Lin:2004en, Zhang:1999bd}. It incorporates the initial deconfined partonic phase, the final hadronic interactions, and the transition between these two phases of matter. AMPT model has four major components as described below.
\begin{itemize}
    \item {\it Initialization}: The initial spatial and momentum distribution of partons, are generated by the \texttt{HIJING} model~\cite{Wang:1991hta}. Multiple scatterings among incoming nucleons are treated in the framework of eikonal formalism. The hard and soft components of particle production are respectively modeled by the formation of energetic minijet partons and soft string excitations. The differential cross-section of the produced mini-jet partons and excited strings are first calculated for pp collisions and are then converted into p+A and A+A collisions by using the in-built Glauber model~\cite{Lin:2004en, Wang:1991hta}. 
    
    \item {\it Transport of partons}: The produced partons are propagated to the Zhang's Parton Cascade (\texttt{ZPC}) model, where the Boltzmann transport equation for partons is solved via cascade method~\cite{Zhang:1997ej}. In the string-melting version of AMPT, all excited strings are converted into hadrons and then decomposed further into their constituent quarks~\cite{He:2017tla}. These partons are then combined with the mini-jet partons and evolve through the \texttt{ZPC} model via two-body elastic parton scatterings~\cite{Lin:2004en}.
    
    \item {\it Hadronization}: The transported partons are then hadronized using either the default mode or the string melting mode of AMPT. In the default mode, the transported partons are combined with their parent strings via the Lund string fragmentation model and then the strings get converted into hadrons~\cite{Andersson:1983ia}. In the string melting mode, transported partons are combined to form hadrons through a quark coalescence mechanism~\cite{Lin:2001zk, He:2017tla}. In this model, a quark and an anti-quark sharing a close phase-space combine to form a meson while the three nearest quarks combine into baryons~\cite{Lin:2004en, Lin:2001zk}.
    
    \item {\it Hadron transport}: Evolution of the resulting hadronic matter is described via hadron cascade, which is based on a relativistic transport (\texttt{ART}) model~\cite{Li:2001xh,Li:1995pra}. The produced hadrons go through a final evolution via baryon-baryon, meson-baryon, and meson-meson scatterings and decays~\cite{Lin:2004en}. 
\end{itemize}

As the quark coalescence mechanism for hadronization implemented in the string melting version of AMPT explains the particle $p_{\rm T}$ spectra and flow at intermediate-$p_{\rm T}$, we use the string melting mode of AMPT (version 2.26t9b) for this study~\cite{Das:2022lqh, Altmann:2024icx, Mallick:2020ium, Mallick:2021hcs, Behera:2021zhi, Mallick:2022alr, Prasad:2022zbr, Behera:2023nwj, Mallick:2023vgi}. For this study, in AMPT, we take the partonic scattering cross-section to be $\sigma_{\rm gg}= 3$~mb and the value of strong coupling constant as $\alpha_s = 0.33$. The Lund symmetric splitting function parameters are set as $a= 0.3$ and $b=0.15$~\cite{Lim:2018huo}. The parton screening mass in the \texttt{ZPC} model is fixed to $\mu = 2.265$~fm$^{-1}$. With the above-mentioned settings in AMPT, we simulate p--O and p--C collisions at $\sqrt{s_{\rm NN}} = 9.9$~TeV.

Since the impact parameter ($b$), the number of participants ($N_{\rm part}$), and the number of binary collisions ($N_{\rm coll}$) cannot be directly measured in experiments, we resort to Glauber model estimations for the same~\cite{Loizides:2017ack, Loizides:TGlauberMC}. Using the publicly available MC Glauber code (\texttt{TGlauberMC$-3.2$})~\cite{Loizides:TGlauberMC}, the impact parameter distribution is sliced to obtain the required centralities of collision. We have also modified the in-built \texttt{HIJING} model in AMPT to incorporate SOG and $\alpha-$cluster nuclear density profiles for both $^{16}\rm O$ and $^{12} \rm C$ nuclei. Due to the unavailability of experimental data for p--O and p--C collisions at the LHC, we tune the AMPT model parameters by comparing them with that of the p--Pb system. A detailed description of the parameter tuning is provided in the Appendix~\ref{Appendix-II}.

\subsection{Two-particle cumulants}
\label{qcumulant}
Azimuthal anisotropy is one of the key observables that can characterize the medium formed in relativistic heavy-ion collisions. The asymmetric pressure gradient created in the medium due to the initial spatial anisotropy of the nuclear overlap region can transform into the final-state momentum space azimuthal anisotropy. This is also known as the anisotropic flow. The anisotropic flow can be quantified as the coefficients of Fourier expansion of the azimuthal momentum distribution of the final-state particles as follows~\cite{ALICE:2014wao},
\begin{equation}
\frac{dN}{d\phi}=\frac{1}{2\pi}\Big(1+\sum_{n=1}^{\infty}2v_{n}\cos[n(\phi-\psi_{n})]\Big).
\label{eq:fourierexpansion}
\end{equation}
Here, $\phi$ is the azimuthal angle, $\psi_n$ is the $n$th harmonic event-plane angle, and $v_{n}$ characterizes the $n$th-order anisotropic flow coefficient. $v_{1}$ stands for the directed flow, $v_{2}$ for elliptic flow and $v_{3}$ characterises the triangular flow, and so on. One can estimate the anisotropic flow of different orders using the following expression~\cite{ALICE:2014wao},
\begin{equation}
    v_{n}=\langle\cos[n(\phi-\psi_{n})]\rangle.
    \label{eq:vneventplane}
\end{equation}
Here, $\langle\dots\rangle$ represents the average over all particles in an event. The estimation of anisotropic flow coefficients from Eq.~\eqref{eq:vneventplane} requires $\psi_n$ whose measurement is not trivial in experiments. In addition, the flow coefficients estimated using Eq.~\eqref{eq:vneventplane} are prone to non-flow effects, such as contributions from jets, and short-range resonance decays. These effects are more pronounced in small collision systems. Thus, to avoid these issues, multi-particle cumulant method is prescribed to estimate the anisotropic flow coefficients. In this method, one does not require the information of $\psi_n$ and the non-flow effects can also be significantly reduced by implementing a proper relative pseudorapidity ($\Delta\eta$) cut between the particles. 

In this study, the anisotropic flow coefficients are estimated using a two-particle $Q$-cumulant method. In this method, the two-particle azimuthal correlations are expressed in terms of a $Q$-vector as follows~\cite{Bilandzic:2010jr, Zhou:2014bba, Zhou:2015iba},
\begin{equation}
Q_{n} = \sum_{j=1}^{M}e^{in\phi_{j}}.
\end{equation}
Here, $M$ is the multiplicity of the event, and $\phi_{j}$ is the azimuthal angle of $j$th hadron. The index `$j$' runs over all the charged hadrons. 
The single-event average two-particle azimuthal correlations are estimated using the following expression,
\begin{equation}
\begin{aligned}
\langle 2 \rangle & = \frac{|Q_{n}|^{2} - M} {M(M-1)}.
\end{aligned}
\label{Eq:Mean24}
\end{equation}

Using Eq.~\eqref{Eq:Mean24}, one can estimate the two-particle cumulants as follows,
\begin{equation}
\begin{aligned}
c_{n}\{2\} & = \langle \langle 2 \rangle \rangle = \frac{\sum_{i=1}^{N_{\rm ev}}(W_{\langle 2 \rangle})_{i}\langle 2 \rangle_{i}}{\sum_{i=1}^{N_{\rm ev}}(W_{\langle 2 \rangle})_{i}}.
\end{aligned}
\label{Eq:c24}
\end{equation}
Here, $\langle\langle~\rangle\rangle$ denotes the average over all particles over all the events. $N_{\rm ev}$ is the total number of events used for the calculations, and $(W_{\langle 2 \rangle})_{i}$ is the weight factor for the $i$th event which takes into account the number of different two-particle combinations in an event with multiplicity $M$. $W_{\langle 2 \rangle}$ can be estimated as follows,
\begin{equation}
    W_{\langle 2 \rangle}=M(M-1).
\end{equation}
One can obtain the value of event-averaged reference flow with the two-particle cumulants using the following expression,
\begin{equation}
    v_{n}\{2\}=\sqrt{c_{n}\{2\}}.
    \label{eq:refvn}
\end{equation}
However, to estimate the differential flow of the Particles Of Interest (POIs), one can define $p_{n}$ and $q_{n}$ vectors with specific kinematic cuts as follows,
\begin{equation}
\begin{aligned}
p_{n}& = \sum_{j=1}^{m_{p}} e^{in\phi_{j}},\\
q_{n}& = \sum_{j=1}^{m_{q}} e^{in\phi_{j}},
\label{Eqpvector}
\end{aligned}
\end{equation}
where, $m_{p}$ is the total number of particles labeled as POIs, and $m_{q}$ is the total number of particles tagged both as reference flow particles (RFP) and POI. One can estimate the single-event averaged differential two-particle azimuthal correlation using the following expression,
\begin{equation}
\begin{aligned}
\langle 2^{'} \rangle &= \frac{p_{n} Q_{n}^{*} - m_{q}} {m_{p}M - m_{q}}.
\end{aligned}
\label{Eq:Mean24p}
\end{equation}
Finally, the differential flow, $v_{2}(p_{\rm T})$, can be estimated using the following equation,
\begin{equation}
\begin{aligned}
v_{n}\{2\}(p_{\rm T}) & = \frac{d_{n}\{2\}}{\sqrt{c_{n}\{2\}}}.
\end{aligned}
\label{eq:vndiff}
\end{equation}
Here, $d_{n}\{2\}$ is the differential $n^{\rm th}$-order cumulant given as,
\begin{equation}
    d_{n}\{2\}=\langle\langle 2^{'} \rangle\rangle=\frac{\sum_{i=1}^{N_{\rm ev}}(w_{\langle 2^{'} \rangle})_{i}\langle 2^{'} \rangle_{i}}{\sum_{i=1}^{N_{\rm ev}}(w_{\langle 2^{'} \rangle})_{i}}.
\end{equation}
The weight factor, $w_{\langle 2^{'} \rangle}$, is given by,
\begin{equation}
    w_{\langle 2^{'} \rangle}=m_{p}M - m_{q}.
\end{equation}

Unfortunately, the $v_{n}(p_{\rm T})$ obtained from Eq.~\eqref{eq:vndiff} possesses contributions from non-flow effects which can be suppressed by appropriate kinematic cuts. For example, one can introduce a pseudorapidity gap between the particles in the two-particle $Q$-cumulant method~\cite{Zhou:2014bba}. Consequently, the whole event is divided into two sub-events, $A$ and $B$, which are separated by a $|\Delta\eta|$ gap. This modifies Eq.~(\ref{Eq:Mean24}) as,
\begin{equation}
\langle 2 \rangle _{\Delta \eta} = \frac{Q_{n}^{A} \cdot Q_{n}^{B *}} {M_{A} \cdot M_{B}}.
\label{Eq:Mean2Gap}
\end{equation}
Here, $Q_{n}^{A}$ and $Q_{n}^{B}$ are the flow vectors from sub-events $A$ and $B$, respectively. $M_{A}$ and $M_{B}$ are the multiplicities corresponding to the sub-events $A$ and $B$, respectively.

The two-particle $Q$-cumulant with a $|\Delta\eta|$ gap is given by,
\begin{equation}
c_{n}\{2, |\Delta\eta|\} = \langle \langle 2 \rangle \rangle _{\Delta\eta} 
\label{Eq:v22Gap10}
\end{equation}
where, one can estimate $\langle \langle 2 \rangle \rangle _{\Delta\eta}$ by using $\langle 2 \rangle _{\Delta \eta}$ from Eq.~\eqref{Eq:Mean2Gap} with $M_{A} \cdot M_{B}$ as the event weights. If we select RFP from one sub-event and POIs from another to estimate the differential flow with a pseudorapidity gap, there is no overlap between POIs and RFP. This modifies Eq.~\eqref{Eq:Mean24p} as,
\begin{equation}
\langle 2^{'} \rangle_{\Delta \eta}  = \frac{p_{n,A} Q_{n,B}^{*} } {m_{p,A}M_{B}}.
\label{Eqmean2pGap}
\end{equation}
Now, one can obtain the two-particle differential cumulant by taking an event average over $\langle 2^{'} \rangle_{\Delta \eta}$ from  Eq.~\eqref{Eqmean2pGap} with an event weight factor $m_{p,A}M_{B}$. Thus, the two-particle differential cumulant is given by, 
\begin{equation}
d_{n}\{2, |\Delta\eta|\}  = \langle \langle 2^{'} \rangle \rangle_{\Delta\eta}.
\label{dn2Gap}
\end{equation}
Finally, the two-particle differential flow coefficient can be estimated using the following equation,
\begin{equation}
v_{n}\{2, |\Delta\eta| \}(p_{\rm T})  = \frac{d_{n}\{2, |\Delta\eta|\}}{\sqrt{c_{n}\{2,|\Delta\eta|\}}}.
\label{vn2EtaGap}
\end{equation}

To estimate the anisotropic flow coefficients, the multi-particle $Q$-cumulants method is adopted in many experiments~\cite{Zhou:2014bba, Li:2014nka, ATLAS:2017hap}. In this study, $p_{\rm T}$ differential second- and third-order flow coefficients, $v_2$ and $v_3$, are estimated using the above equations by setting the $n=$ 2 and 3, respectively. To estimate the anisotropic flow coefficients, we use all the charged hadrons within the pseudorapidity region, $|\eta_{\rm lab}|<2.5$. The RFP are charged hadrons selected within $|\eta_{\rm lab}|<2.5$ and $0.2<p_{\rm T}<5.0$ GeV/c. In addition, to reduce the non-flow effects from the two-particle $Q$-cumulant method, we use a pseudorapidity gap, $|\Delta\eta|>1.0$, in the two-subevent method.

\section{}
\label{Appendix-II}

\begin{figure}
    \centering
    \includegraphics[scale=0.43]{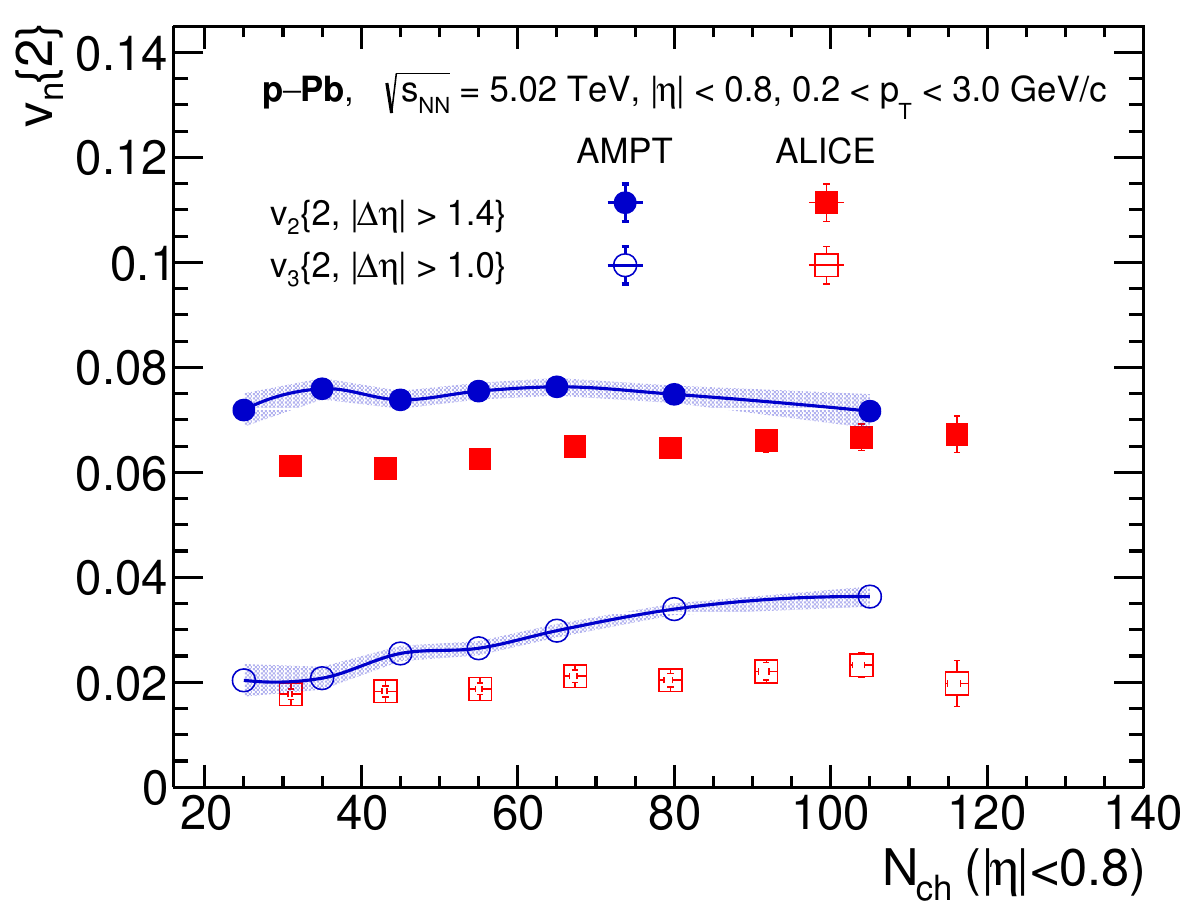}
    \caption{(Color Online) Multiplicity dependence of $v_{2}\{2\}$ and $v_{3}\{2\}$ predicted from AMPT and from ALICE measurements~\cite{ALICE:2019zfl} for p--Pb collisions at $\sqrt{s_{\rm NN}} = 5.02$ TeV.}
    \label{fig:v3v3datacomp}
\end{figure}

$v_{2}\{2\}$ and $v_{3}\{2\}$ are estimated as a function of mid-rapidity multiplicity ($N_{\rm ch} (|\eta|<0.8)$) in p--Pb collisions at $\sqrt{s_{\rm NN}}=~5.02$~TeV from AMPT and are compared to the ALICE measurements in Fig.~\ref{fig:v3v3datacomp}. 
Similar settings of the AMPT parameters are kept as discussed in Appendix~\ref{Appendix-I}, to simulate p--Pb collisions.
From the figure, it appears that AMPT in string melting mode slightly overestimates the values of $v_{2}\{2\}$ and $v_{3}\{2\}$ as a function of $N_{\rm ch}$ in p--Pb collisions at $\sqrt{s_{\rm NN}}=5.02$~TeV compared to the ALICE measurements. Interestingly, AMPT could qualitatively reproduce a very similar trend that of the ALICE measurements.  



\begin{figure*}
    \centering
    \includegraphics[scale=0.42]{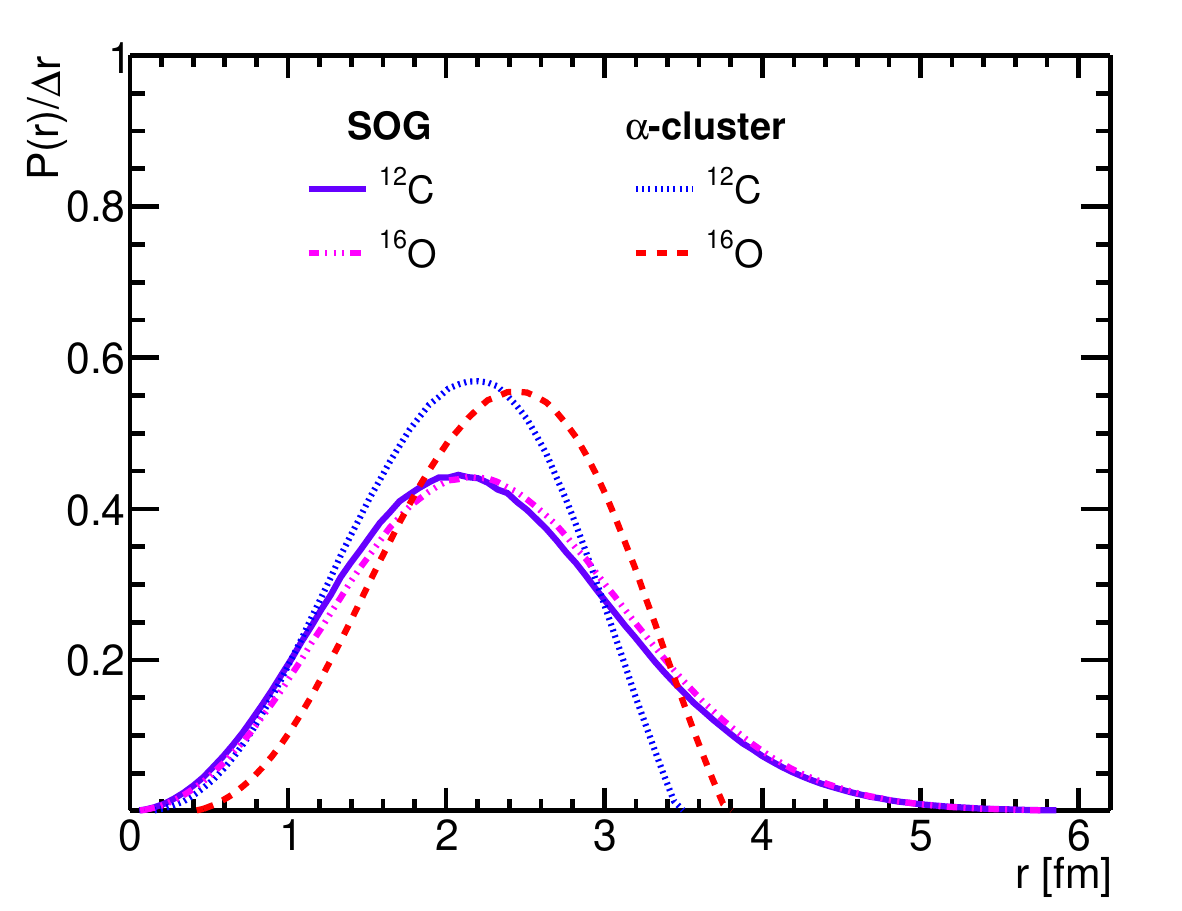}
    \includegraphics[scale=0.42]{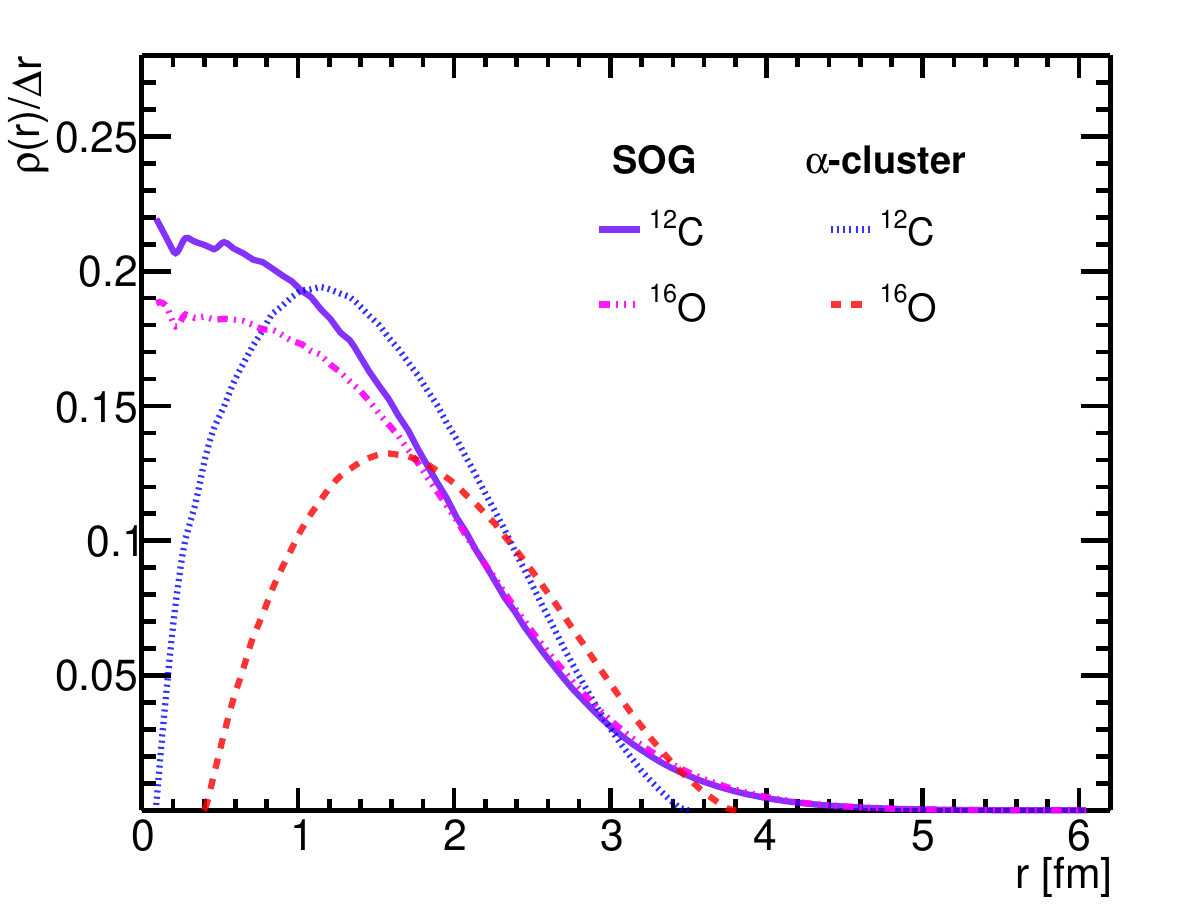}
    \caption{(Color Online) Probability (left) and nuclear density (right) distributions of the radial position of the nucleons inside $^{12}$C and $^{16}$O nucleus considering SOG and $\alpha$-cluster nuclear density profiles.}
    \label{fig:probr}
\end{figure*}




The left panel of Fig.~\ref{fig:probr} shows the probability distribution of the radial position of the nucleons inside $^{12}$C and $^{16}$O nuclei. The nucleon probability distribution considering an $\alpha$-cluster nuclear distribution is compared with the SOG nuclear distribution for both $^{12}$C and $^{16}$O nuclei, where the $\alpha$-cluster nuclear distribution retains a more compact structure as compared to the SOG nuclear density profile. In addition, one finds that the peak positions of the probability distribution for both the nuclear density profiles are slightly lowered for the $^{12}$C case when compared with the $^{16}$O nucleus. The right panel of Fig.~\ref{fig:probr} shows the nuclear density distribution as a function of the distance from the center of the nucleus. The SOG density profile is more dense at the center, while the density of nuclear matter for the clustered structure is largest at a finite distance from the center.

\end{document}